\documentclass[reprint, preprintnumbers]{revtex4-2}

\usepackage{graphicx}
\usepackage{dcolumn}
\usepackage{bm,float,physics,bbold,amssymb, amsfonts, amsmath,braket}
\usepackage{hyperref}
\hypersetup{
  colorlinks   = true,    
  urlcolor     = blue,   
  linkcolor    = blue,   
  citecolor    = blue 
}
\usepackage[mathlines]{lineno}

\usepackage{xcolor}
\usepackage{soul}
\usepackage{lipsum}
\usepackage{mathtools}
\usepackage{cuted}

\begin{document}

\title{A general approach to backaction-evading receivers with magnetomechanical and electromechanical sensors}

\author{Brittany Richman$^{1,2,3}$}
\email{brr215@umd.edu}
\author{Sohitri Ghosh$^{1,2,3,9}$}
\author{Daniel Carney$^5$}
\author{Gerard Higgins$^{7,8}$}
\author{Peter Shawhan$^{2}$}
\author{C. J. Lobb$^{1,2,6}$}
\author{Jacob M. Taylor$^{1,3,4}$}
\affiliation{$^1$Joint Quantum Institute (JQI), College Park, Maryland 20742, USA\\
$^2$Department of Physics, University of Maryland, College Park, Maryland 20742, USA\\
$^3$Joint Center for Quantum Information and Computer Science (QuICS), College Park, Maryland 20742, USA\\
$^4$National Institute of Standards and Technology (NIST), Gaithersburg, Maryland 20899, USA\\
$^5$Physics Division, Lawrence Berkeley National Laboratory, Berkeley, California 94720-8153, USA\\
$^6$Maryland Quantum Materials Center (QMC), College Park, Maryland 20742, USA\\
$^7$Institute for Quantum Optics and Quantum Information (IQOQI), Austrian Academy of Sciences, A-1090 Vienna, Austria \\
$^8$Department of Microtechnology and Nanoscience (MC2), Chalmers University of Technology, SE-412 96 Gothenburg, Sweden \\
$^9$Theory Division, Fermi National Accelerator Laboratory, Batavia, IL 60510, USA
}

\date{\today}

\begin{abstract}
Today's mechanical sensors are capable of detecting extremely weak perturbations while operating near the standard quantum limit. However, further improvements can be made in both sensitivity and bandwidth when we reduce the noise originating from the process of measurement itself --- the quantum-mechanical backaction of measurement --- and go below this `standard' limit, possibly approaching the Heisenberg limit. One of the ways to eliminate this noise is by measuring a quantum nondemolition variable such as the momentum in a free-particle system. Here, we propose and characterize theoretical models for direct velocity measurement that utilize traditional electric and magnetic transducer designs to generate a signal while enabling this backaction evasion.
We consider the general readout of this signal via electric or magnetic field sensing by creating toy models analogous to the standard optomechanical position-sensing problem, thereby facilitating the assessment of measurement-added noise.
Using simple models that characterize a wide range of transducers, we find that the choice of readout scheme --- voltage or current --- for each mechanical detector configuration implies access to either the position or velocity of the mechanical sub-system. This in turn suggests a path forward for key fundamental physics experiments such as the direct detection of dark matter particles. 
\end{abstract}

\preprint{FERMILAB-PUB-23-700-T}

\maketitle

\section{Introduction}
\label{intro}

The ability of the state-of-the-art quantum sensors to monitor the position of objects with high precision~\cite{caves1980measurement,BAEdemo7} has driven tremendous advances in fundamental physics, particularly in the first detection of gravitational waves~\cite{firstLIGO}. Recently there has been renewed interest in the use of momentum measurement~\cite{purdue2002practical,danilishin2019advanced} for ultra-sensitive force detection, specifically for the purposes of dark matter detection~\cite{sohitri}. Approaches for monitoring or measuring particulate dark matter by observing changes in the momentum of test particles, such as those being considered by the Windchime collaboration~\cite{carney2020proposal,snowmass}, represent a key motivation to explore and develop impulse metrology for broadband force sensing.

The measurement of weak forces generally requires the transduction of the induced motion of a system into an electrical or optical signal. Thus, estimating such forces is limited by both technical issues, such as thermal noise and instrumental noise, but also by the noise limits arising from the act of measurement itself~\cite{caves1980measurement}. This is usually characterized by the standard quantum limit~(SQL), which places a lower bound on how precisely the conjugate variables of a system can be measured.
Recently, efforts to get beyond the SQL with mechanical sensors have yielded substantial successes, in part by reducing or removing the effects of measurement backaction~\cite{BAEdemo1,BAEdemo2,BAEdemo3,BAEdemo4,kimble2001conversion,purdue2002practical,BAEdemo7}. For practical or fundamental applications, including gravitational wave detection~\cite{caves1980measurement,purdue2002practical,kimble2001conversion,gravityBAE} and, more recently, the detection of potential dark matter candidates~\cite{carney2020proposal,sohitri,beckey2023quantum}, the ability to measure beyond the SQL may also be paired with the need to do so over a wide range of signal frequencies, as in the case of broadband signals from a black hole in-spiral or the case of signals from rapidly moving particulate dark matter.

With these interests in mind, a particular opportunity emerges for going beyond the SQL: quantum nondemolition (QND) measurement. The simplest example occurs when measuring a free particle, or a harmonic oscillator well above its resonance frequency: measurements of momentum at different times commute, even when accounting for the evolution of the system, and thus backaction can be pushed into the position variable without it disturbing subsequent measurements, thereby circumventing the SQL~\cite{caves1980measurement, thorne1978quantum}. In practice, the canonical momentum of a \textit{combined} probe and mechanical system is not a QND variable~\cite{danilishin2019advanced}. Nevertheless, measurement of the mechanical sub-system's momentum, rather than its position, can provide a significant reduction of the measurement backaction.

In this work, we explore options to transduce a force signal to an electrical signal, enabling measurement using a parametric cavity in the microwave regime or optical regime (Fig.~\ref{fig:visual}), a typical paradigm from the optomechanics and electromechanics community~\cite{GKmm1,GKmm2,GKmm3,em1,em2,em3}. In contrast to most work in the field, here we are focused on broadband frequency signals (impulses delivered over very short times), typically well above the mechanical resonance (as indicated in Fig.~\ref{fig:visual}e), and thus seek to exploit the QND opportunities such measurements enable. Our detailed examination reveals that the mechanical variable (position or momentum) accessed is dependent on the chosen measurement readout of the electrical system: via charge or flux. This indicates two different approaches for broadband backaction-evading measurement transduced to the microwave domain. Of specific interest in our case is leveraging the intuition of Faraday's law --- that velocity creates a voltage --- to find a magnetomechanical scheme that utilizes a traditional voice coil, which is used in dynamic microphones and loudspeakers~\cite{Crandall,Fitz}. We find QND measurement of velocity occurs when this coil's voltage is measured rapidly. Surprisingly, we also find that for an electromechanical detector, such as a variable-position capacitor used in a condenser microphone~\cite{Crandall,Fitz,Kamm,Bishop}, we also can get backaction evasion by rapid measurement of the current from the microphone.

We remark that there are other electrical measurement approaches, such as DC current or voltage measurement, that can in principle allow direct velocity measurements to be naturally implemented without the use of a parametric cavity. However, there are a variety of technical challenges in achieving the SQL with DC current or voltage measurement.
We also note that compared to the optical domain, electrical systems overall offer a more energy efficient readout by operating in the microwave regime. As a consequence, far less power is required to interrogate a system's sensors and achieve a SQL-level resolution~\cite{lecocq2016mechanically, palomaki2013coherent, regal2008measuring, zeuthen2018electrooptomechanical}.

This paper is organized as follows. In Section~\ref{S1}, we introduce and explore two detector configurations that utilize different electrical transducers: a magnetomechanical detector scheme and an electromechanical detector scheme. The general readout of a transduced electrical signal is considered in Section~\ref{S2}, by using a parametric cavity for either electric field or magnetic field sensing. This creates an optomechcanical analog that enables the use of standard techniques from optomechanics. In Section~\ref{S3} we assess measurement-added noise and consider an example signal, comparing the four cases given by the combination of the two detector schemes presented in Section~\ref{S1} and the two readout options discussed in Section~\ref{S2}. We conclude and discuss the implications of our results in Section~\ref{Conclusions}.

\section{Transducers}
\label{S1}

\subsection{Magnetomechanical Detector Scheme}
\label{s1mm}

\begin{figure*}
\centering
\includegraphics[width=0.96 \textwidth]{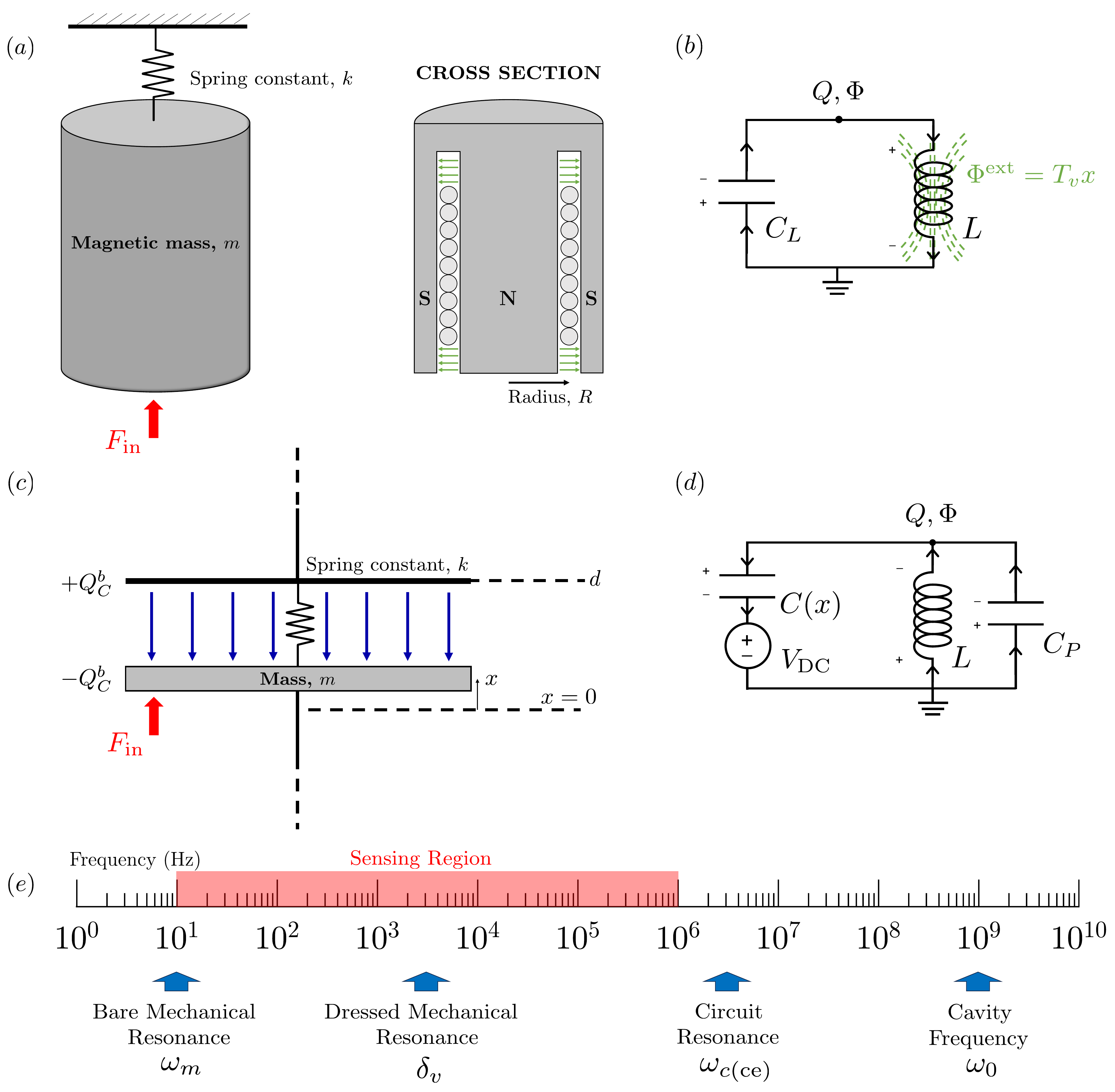}
\caption{(a) Schematic representation of the magnetomechanical transducer, consisting of a structured magnetic mass $m$ suspended by a spring with spring constant $k$. In the cross section, the radial magnetic field in the air gap experienced by the coil is indicated in green. The lumped-element detector circuit for the magnetomechanical case is shown in (b), where the magnetomechanical element of (a) is modeled as an inductor $L$ threaded by an external flux $\Phi^{\text{ext}} = T_v x$ (shown in green) dependent on the position of the magnetic mass. Sign conventions for analysis are indicated as well as the reference ground node and node variables $Q,\Phi$. (c) Schematic representation of the electromechanical transducer, including the electric field (dark blue) between the charged plates of the mechanical capacitor consisting of a movable plate of mass $m$ and charge $-Q^b_C$ connected to a fixed plate of charge $+Q^b_C$ by a spring with spring constant $k$. Equilibrium distances are indicated for reference. In (d), the electromechanical element of (c) is depicted in a lumped-element detector circuit via the capacitance $C(x)$. Sign conventions for analysis are indicated as well as the reference ground node and node variables $Q,\Phi$. In (e), we indicate the relative scale of frequencies in the system for our chosen parameters, including that of the to-be-incorporated parametric cavities (see Fig.~\ref{fig:cavity}). The input force on the test masses is indicated in red, resulting in the broad sensing region indicated by the red shaded region.}
\label{fig:visual}
\end{figure*}

We begin by examining the magnetomechanical transducer and detector scheme shown in Figs.~\ref{fig:visual}a and \ref{fig:visual}b, respectively, which operate as a consequence of Faraday's law~\cite{jackson,zangwill}.
This fundamental principle of electromagnetism describes the voltage generated when a time-varying magnetic flux threads a conducting loop. This allows us to first consider the magnetomechanics of the transducer and establish the transducer constant before deriving the Hamiltonian describing the total detector scheme. A review of Faraday's law and its application in the example we consider can be found in Appendix~\ref{faraday}.

The magnetomechanical transducer we consider, shown in Fig.~\ref{fig:visual}a, consists of two main elements: a magnetic test mass and a pick-up coil. We take the pick-up coil to be superconducting so as to neglect any internal dissipation. The mass is attached to a spring with spring constant $k$. The test mass is a cylindrical magnetic structure of mass $m$ that contains an annular air gap at a radius $R$ through the length of the cylinder, as shown in the cross-section in Fig.~\ref{fig:visual}a. This magnetic mass is arranged such that within the air gap a uniform radial magnetic field $\bold{B} = B_r \bold{\hat{r}}$ is maintained. Embedded within the air gap is a pick-up coil of radius $R$ and turn number~$N$. This is known as the voice-coil configuration, common to dynamic microphones and loudspeakers~\cite{Crandall,Fitz}. This transducer allows for a simple closed-form expression for the induced voltage, which due to the uniform field, is purely velocity-dependent, as we now show.

We consider an impulse which causes the magnetic mass to move with a velocity $\bold{v} = -v \bold{\hat{z}}$. For velocities much less than the speed of light, it is both equivalent and convenient to compute the voltage in the rest frame of the magnet, rather than the rest frame of the voice coil (as in our detection scheme). In the rest frame of the magnet, the voice coil moves with velocity $\bold{v} = v \bold{\hat{z}}$, experiencing the magnet's uniform magnetic field $\bold{B} = B_r \bold{\hat{r}}$ and no electric field ($\bold{E} = 0$). We then use Faraday's law to calculate the induced voltage across the voice coil~to~be:
\begin{equation}
\label{eq:voltage}
\varepsilon = 2 \pi N R B_r v  = T_v v \enspace .
\end{equation}
We note that this induced voltage is proportional to velocity via the transducer constant $T_v = 2 \pi N R B_r$, which is simply the product of the length of wire used in the voice coil and the magnetic field. Through this constant, the mechanical motion of the magnetic test mass is transduced to a voltage, which may be read out electrically.

The magnetomechanical detector scheme, shown in Fig.~\ref{fig:visual}b, is a parallel LC-circuit that models the transducer as an inductor with inductance $L$ threaded by an external flux $\Phi^{\text{ext}}$, due to the interaction between the voice coil and magnetic test mass of the transducer, with the capacitance $C_L$ in parallel. We then employ the usual techniques for circuit quantization~\cite{devoret} to come to a Hamiltonian description of the magnetomechanical detector scheme that includes the mechanical motion. This derivation and its associated details can be found in Appendix~\ref{circuits101}.
The coupling between the mechanics and the circuit occurs via the presence of the position-dependent external flux $\Phi^{\text{ext}} = T_v x$, where we identify $\dot{\Phi}^{\text{ext}}$ with the induced voltage $\varepsilon$ in Eq.~(\ref{eq:voltage}) and $x$ with the position of the test mass.

We first consider a simplified case, treating $\Phi^{\text{ext}}$ as only a time-dependent external flux.
This leads to two distinct gauge descriptions of the circuit that differ in how they couple to the external flux: one to its time-derivative (a voltage) via a term $Q \dot{\Phi}^{\text{ext}}$ and the other to the external flux via a term $\Phi \Phi^{\text{ext}}/L$, where $Q$ and $\Phi$ correspond to the circuit's degrees of freedom.
Upon promoting the degrees of freedom to quantum operators $\hat{Q}$ and $\hat{\Phi}$, which satisfy the commutation relation $[\hat{Q},\hat{\Phi}] = -i \hbar$, we see that these two equivalent descriptions are related by a time-dependent gauge transformation given by the unitary:
\begin{equation}
\label{eq:GT}
\hat{U} = e^{-i \hat{Q} \Phi^{\text{ext}}/\hbar} \enspace .
\end{equation}
We note that recent work \cite{you,youexp,DV,riwar} has focused on the term $Q \dot{\Phi}^{\text{ext}}$ of the first gauge, exploring other equivalent descriptions that eliminate its necessity. In contrast, we focus on both of these gauge choices to highlight and compare how the two gauges couple the circuit to the mechanical degrees of freedom.

Incorporating the mechanical degrees of freedom introduces the canonical momentum $p$ into the Hamiltonian. Importantly, $p$ is not necessarily the mechanical momentum $m \dot{x}$; the gauge choice determines whether or not this is the case. In particular, in the first gauge, $p = m \dot{x} + C_L T_v (\dot{\Phi} + T_v \dot{x})$. In the second gauge, $p = m \dot{x}$.
As before, upon promoting the degrees of freedom to operators where $[\hat{Q},\hat{\Phi}] = -i \hbar$ and $[\hat{p},\hat{x}] = -i \hbar$, we find these two equivalent descriptions are now related via a more general unitary (gauge) transformation
\begin{equation}
\label{eq:GTmech}
\hat{U} = e^{-i T_v \hat{Q} \hat{x}/\hbar} \enspace .
\end{equation}
Altogether, we find the Hamiltonians
\begin{equation}
\label{eq:H1mV}
\begin{split}
\hat{H}_1^{(E,v)} = \frac{\hat{p}^2}{2m}& + \frac{1}{2} k \hat{x}^2 - \frac{T_v}{m}\hat{Q} \hat{p} \\
& + \frac{\hat{Q}^2}{2}\left(\frac{1}{C_L} + \frac{T_v^2}{m}\right) + \frac{\hat{\Phi}^2}{2L} 
\end{split}
\end{equation}
and
\begin{equation}
\label{eq:H2mV}
\hat{H}_2^{(E,v)} = \frac{\hat{p}^2}{2m} + \frac{1}{2} k \hat{x}^2 + \frac{\hat{Q}^2}{2 C_L} + \frac{(\hat{\Phi} - T_v \hat{x})^2}{2L} \enspace ,
\end{equation}
where $\hat{Q}$ and $\hat{p}$ are the canonical node charge and momentum conjugate to the node flux $\hat{\Phi}$ and position $\hat{x}$, respectively. With only the capacitance $C_L$ connected to the node, $\hat{Q}$ directly corresponds to the charge on this capacitor's plates. The subscripts in Eqs.~(\ref{eq:H1mV}) and (\ref{eq:H2mV}) enumerate the two gauge choices in this detector scheme, while the superscripts indicate the readout scheme (explored in Section~\ref{S2}) and detector configuration.

Examination of Eqs.~(\ref{eq:H1mV}) and (\ref{eq:H2mV}) reveals an interesting feature of this detector configuration: equivalent descriptions contain different couplings to the mechanical degrees of freedom. In the first gauge, the voice coil and its mechanics are coupled through momentum and charge, while in the second gauge, position and flux are coupled. We account for both of these gauge descriptions in the magnetomechanical case as we consider the different readout schemes in Section~\ref{S2}. However, these different gauges are equivalent descriptions and do not yield any differences in performance, as discussed in Section~\ref{S3}.

\subsection{Electromechanical Detector Scheme}
\label{s1em}

In contrast to the magnetomechanical configuration, an electromechanical detection scheme is governed by electrostatic principles. In this case, we begin with the Hamiltonian description and leverage this understanding to establish how mechanical motion translates to an electrical signal. We consider the electromechanical transducer and detector scheme shown in Figs.~\ref{fig:visual}c and \ref{fig:visual}d, respectively. The electromechanical transducer consists of two oppositely charged plates with charge $\pm Q^b_C$ connected by a spring with spring constant~$k$. This forms a capacitor with one fixed plate and one movable plate of mass $m$ whose capacitance is a function of the position of the movable plate, namely, $C(x) = \frac{\epsilon_0 A}{d_0-x}$. We take the area of the plates $A$ to be much larger than their uncharged equilibrium separation $d_0$. As the massive plate moves due to an impulse, the capacitance changes, thereby altering the charge on the plates and the voltage across them. In the language of fields, as the plate moves, the uniform electric field between the plates changes strength. These are the working principles behind condenser microphones~\cite{Crandall,Fitz,Kamm,Bishop}.

We consider such a mechanically-varying capacitor in the detector circuit shown in Fig.~\ref{fig:visual}d. In addition to the capacitance $C(x)$, this circuit consists of a voltage source $V_{\text{DC}}$ in series to charge the capacitor, in parallel with a large inductor $L$ and its parasitic capacitance $C_P$. We then employ the usual techniques for circuit quantization~\cite{devoret} to come to a Hamiltonian description of the detector circuit.
Due to the inverse dependence on position in the capacitance $C(x)$, the coupling between the circuit and mechanical degrees of freedom is nontrivial in form. However, we can linearize the coupling by considering impulses that amount to only small displacements from equilibrium. Details of this derivation can be found in Appendix~\ref{circuits101}.

In total, we come to a linearized Hamiltonian of the form
\begin{equation}
\label{eq:Hem}
\begin{split}
\hat{H}^{(E,x)} = \frac{\hat{p}^2}{2m} + \frac{1}{2} &k_{\text{eff}} (\hat{x}-x_0)^2 - \frac{T_x}{C_P} (\hat{Q}-Q_0) (\hat{x}-x_0) \\
& + \frac{(\hat{Q}-Q_0)^2}{2 C_{\text{eff}}} + \frac{\hat{\Phi}^2}{2L} + V(Q_0,x_0) \enspace ,
\end{split}
\end{equation}
where $\hat{Q}$ and $\hat{p}$ are the canonical node charge and canonical momentum, which are conjugate to the node flux $\hat{\Phi}$ and position $\hat{x}$, respectively. In this case, $\hat{Q}$ corresponds to the sum of the charge on the plates of the two capacitors connected to the node, while $\hat{p}$ is the mechanical momentum of the movable plate. We define the effective capacitance as $C_{\text{eff}} = C(x_0) + C_P$ and the effective spring constant to be $k_{\text{eff}} = k - \frac{C_{\text{eff}} T_x^2}{C(x_0) C_P}$. The point~$\lbrace Q_0, x_0 \rbrace$ corresponds to the equilibrium node charge and position when the plates are charged. While the distance~$d_0$ corresponds to the separation of the plates when uncharged, as the plates charge, the plate separation decreases as electrostatic attraction shifts the position of the movable plate closer to its counterpart. This equilibrium point corresponds to the position where the force of electrostatic attraction and the restorative force of the spring exactly balance, and is a function of the spring constant $k$, the voltage bias $V_{\text{DC}}$, and the geometry of $C(x)$, namely, the area $A$ and plate separation $d_0$. We define the energy associated with this equilibrium charge and position configuration as $V(Q_0,x_0)$, whose functional form is given in Appendix~\ref{circuits101}.

Importantly, this linearization procedure enables us to define the transducer constant $T_x$ in this system:
\begin{equation}
\label{eq:TE}
T_x = C_P \frac{\left( Q_0-C_P V_{\text{DC}} \right)}{\epsilon_0 A} \left( \frac{C(x_0)}{C_{\text{eff}}} \right)^2 \enspace ,
\end{equation}
which characterizes how changes in the position $x$ of the movable plate result in changes of the charge on the mechanically-varying capacitor's plates $\pm Q^b_C$, i.e.,
\begin{equation}
\label{eq:TErelx}
Q^b_C \approx  \frac{C(x_0)}{C_{\text{eff}}} (Q - C_P V_{\text{DC}}) + T_x (x-x_0) \enspace .
\end{equation}
With this description, it is clear that the charge on the plates of the capacitor is sensitive to the position of the movable plate while the current through the mechanically-varying capacitor is sensitive to the velocity:
\begin{equation}
\label{eq:TErelv}
i^b_C = \dot{Q}^b_C \approx  \frac{C(x_0)}{C_{\text{eff}}} \dot{Q} + T_x \dot{x} \enspace .
\end{equation}
From Eqs.~(\ref{eq:TErelx}) and (\ref{eq:TErelv}), we see how the position of the movable plate affects the charge (and by extension, the voltage) on the mechanically-varying capacitor and how the motion of the plate results in a current. Together with Eq.~(\ref{eq:Hem}), these equations characterize the behavior and response of the electromechanical transducer shown in Fig.~\ref{fig:visual}c.

Up until this point, we have focused on the detector configurations and the corresponding electrical signals produced as a result of the motion of a test mass. In the following section, we shift our attention to the measurement of this signal by considering various readout schemes.

\section{Idealized receivers}
\label{S2}

\begin{figure*}
\centering
\includegraphics[width=0.96 \textwidth]{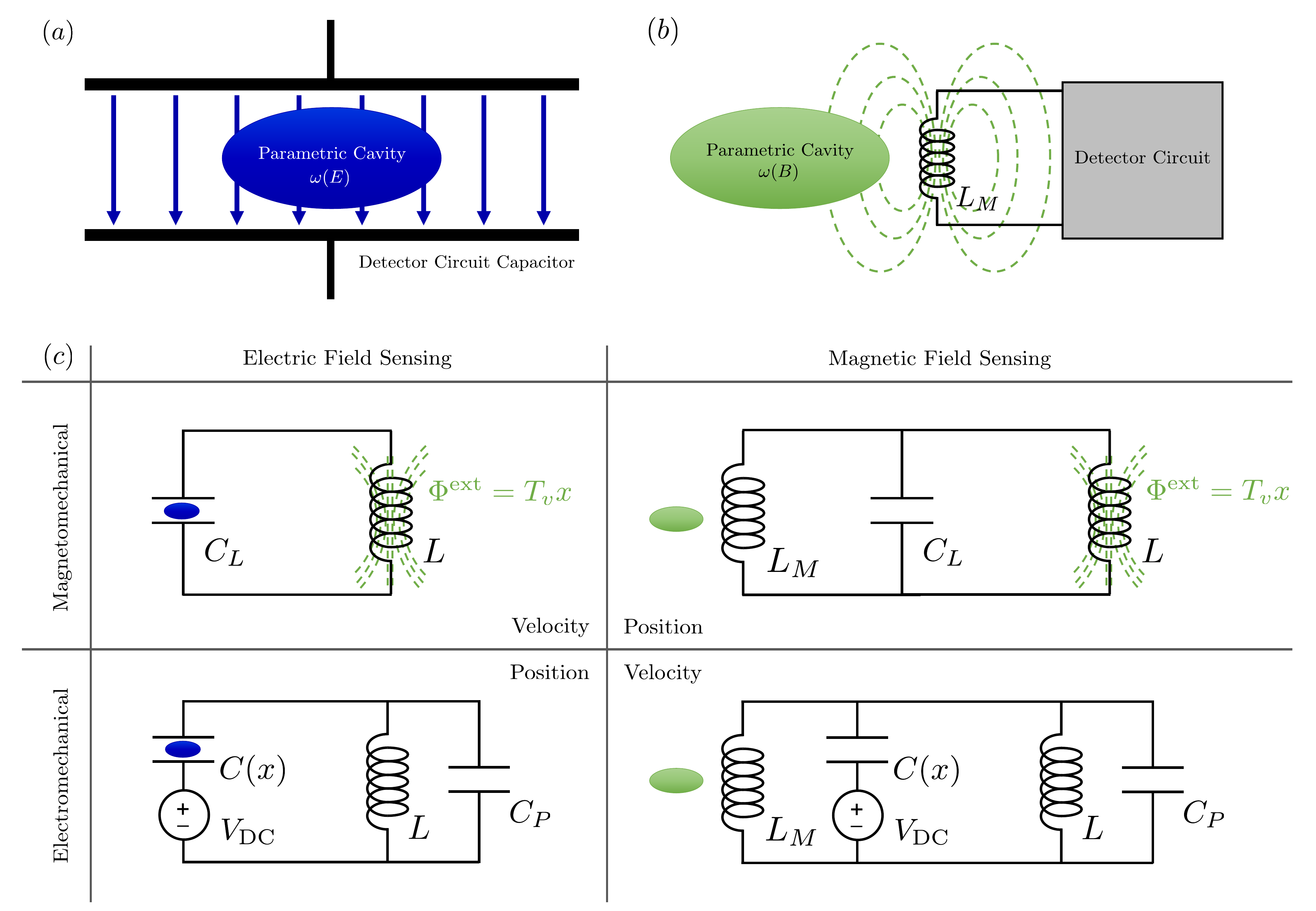}
\caption{(a) Schematic representation of the configuration for voltage measurement via electric field sensing using a parametric cavity (such as a rf-SET) with a resonance frequency that depends on the electric field $E$ (indicated in dark blue) inside a capacitor of each detector circuit. (b) Schematic representation of the configuration for current measurement via magnetic field sensing using a parametric cavity (such as a resonator terminated with a DC SQUID) with a resonance frequency that depends on the magnetic field $B$ (indicated in green) generated by an inductor $L_M$ connected to each detector circuit. We show here that the parametric cavity readout depends on the electric field or the magnetic field and is naturally gauge invariant. In (c), we provide a visual table showing the different combinations of measurement schemes and detector configurations considered in the main text, including small blue and green ovals representing the cavities to indicate which circuit elements interact with the parametric cavities in each case as well as a labelling of which schemes access the mechanical position or velocity.}
\label{fig:cavity}
\end{figure*}

Our interest lies in the measurement-added noise associated with each detector scheme, in which a mechanical signal of interest is transduced to an electrical one. In this section, we consider readout schemes to access the mechanical degrees of freedom via the detector circuits' degrees of freedom. Because measurement-added noise in optomechanical systems is well understood, we imagine reading out the degrees of freedom of the detector circuits using a parametric cavity whose frequency depends on the electric or magnetic fields generated in each detector circuit, as shown in Fig.~\ref{fig:cavity}. This parametric cavity approach is exemplified by the rf-SET (radio-frequency single-electron transistor)~\cite{rfSET}, a Cooper-pair box connected to a resonant circuit where variations in the local electric field change the circuit properties and can be detected in reflectometry. Thus our approach amounts to a measurement of the mechanically-generated electrical signal as a voltage via the electric field of a capacitor or a current via the magnetic field of an inductor. As such, we consider voltage and current readout schemes for both the magnetomechanical and electromechanical detection schemes of Section~\ref{S1}, including the two gauge descriptions outlined in Section~\ref{s1mm}.

We note again that in practice, there are methods for reading out the degrees of freedom of an electrical circuit which may be more direct, such as using a Cooper-pair box or a superconducting quantum interference device (SQUID)~\cite{CPBex,CPBex2,CPBex3,CPBex4,gerard,gerard2,SQ1,SQog,SQ2}. However, this Gedankenexperiment serves as a useful scaffolding for us to develop our understanding of measurement-added noise in electrical systems.

\subsection{Electric field sensing}
\label{ss2vm}

To access a detector circuit's voltage, we consider measuring the electric field across a capacitor using a parametric cavity sensitive to electric fields, as indicated in Fig.~\ref{fig:cavity}a and the left column of Fig.~\ref{fig:cavity}c, and exemplified in a rf-SET~\cite{rfSET} or similar device from circuit quantum electrodynamics (circuit QED)~\cite{CPBex,CPBex2,CPBex3,CPBex4} in the microwave domain.
We then imagine the parametric cavity to have a resonance frequency that depends on the electric field of a capacitor coupled to the cavity.
For a parallel plate capacitor, we express the electric field in terms of the charge on the plates $Q^b$ and their area $A$, or in terms of the voltage across the plates $v^b$ and their separation~$d^\prime$: $E = \frac{Q^b}{\epsilon_0 A} = \frac{v^b}{d^\prime}$.
Here, $Q^b$ and $v^b$ refer to the branch charge and branch voltage of the capacitor, which are unique to each detector configuration.
We use the superscript $b$ to denote branch variables throughout the text; this is to distinguish from other variables, such as velocity $v$, present in our analysis. Branch variables and their role in circuit analysis are discussed in Appendix~\ref{circuits101}.

In the magnetomechanical detector configuration, we take the parallel capacitance $C_L$ as the capacitor coupled to the parametric cavity. As described in Section~\ref{s1mm}, the charge on this capacitor's plates corresponds to the node charge $Q$. One can also confirm, using Appendix~\ref{circuits101}, that the voltage across the capacitor, $v^b_{C_L} = \dot{\Phi}^b_{C_L} = -Q/C_L$, is gauge-independent.

Therefore, in both gauges we express the electric field in terms of the degrees of freedom of the circuit as
\begin{equation}
\label{eq:Efield}
E^v = -\frac{Q}{\epsilon_0 A} = -\frac{Q}{C_L d_L} \enspace ,
\end{equation}
where we have taken $C_L = \frac{\epsilon_0 A}{d_L}$ with $d_L$ the separation of the plates.

In the electromechanical case, we take the mechanically-varying capacitor $C(x)$ to be the capacitor coupled to the parametric cavity. To express the electric field in this capacitor in terms of the degrees of freedom of the circuit, we use the linearized expression for the charge on the plates of the capacitor $C(x)$, given by Eq.~(\ref{eq:TErelx}). This yields an approximate expression for the electric field of the form
\begin{equation}
\label{eq:emEfield}
E^x \approx  \frac{C(x_0)}{C_{\text{eff}}} \frac{(Q - C_P V_{\text{DC}})}{\epsilon_0 A} + \frac{T_x}{\epsilon_0 A} (x-x_0) \enspace ,
\end{equation}
and ensures a linear coupling of position to the cavity for small displacements.

We then incorporate the parametric cavity into the Hamiltonian description from Section~\ref{S1}, given by Eqs.~(\ref{eq:H1mV}), (\ref{eq:H2mV}), and (\ref{eq:Hem}). The cavity Hamiltonian takes the usual form,
\begin{equation}
\label{eq:Ecavity}
\hat{H}_{\text{cav}}^E = \hbar \omega(E) \hat{a}^\dagger \hat{a} \enspace ,
\end{equation}
where $\hat{a}^\dagger,\hat{a}$ are the creation and annihilation operators of the cavity mode that satisfy the commutation relation $[\hat{a}, \hat{a}^\dagger]=1$ and $\omega(E)$ is the resonance frequency of the cavity that depends on the electric field $E$. To generate the usual optomechanical coupling, we expand $\omega(E)$ about some equilibrium field value $E_0$:
\begin{equation}
\label{eq:genEwexp}
\omega(E) = \omega(E_0) + \eta_E (E-E_0) + ... \enspace ,
\end{equation}
where we define the cavity's sensitivity to electric fields $\eta_E = \frac{d \omega(E)}{dE}\rvert_{E=E_0}$.
We then make the substitutions for $E^v$ and $E^x$ from Eqs.~(\ref{eq:Efield}) and (\ref{eq:emEfield}) and truncate at linear order to define the cavity frequencies
\begin{equation}
\omega(E^v) = \omega_0^{(E,v)} - g_Q^{(E,v)} \hat{Q}
\end{equation}
and
\begin{equation}
\omega(E^x) = \omega_0^{(E,x)} + g_Q^{(E,x)} \hat{Q} + g_x \hat{x} \enspace ,
\end{equation}
where we define the coupling constants $g_Q^{(E,v)} = \frac{\eta_E}{C_L d_L}$, $g_Q^{(E,x)} = \frac{\eta_E C(x_0)}{C_{\text{eff}} \epsilon_0 A}$, and $g_x = \frac{\eta_E T_x}{\epsilon_0 A}$. We have also defined the rescaled cavity frequencies $\omega_0^{(E,v)} = \omega(E_0^v) - \eta_E E_0^v$ and $\omega_0^{(E,x)} = \omega(E_0^x) - \eta_E E_0^x  - g_Q^{(E,x)} C_P V_{\text{DC}} - g_x x_0$.

Altogether, we write the cavity Hamiltonian in Eq.~(\ref{eq:Ecavity}) for the magnetomechanical and electromechanical detector configurations as 
\begin{equation}
\label{eq:vcavmm}
\hat{H}_{\text{cav}}^{(E,v)} = \hbar \omega_0^{(E,v)} \hat{a}^\dagger \hat{a} - \hbar g_Q^{(E,v)} \hat{Q} \hat{a}^\dagger \hat{a} 
\end{equation}
and
\begin{equation}
\label{eq:vcavem}
\hat{H}_{\text{cav}}^{(E,x)} = \hbar \omega_0^{(E,x)} \hat{a}^\dagger \hat{a} + \hbar \left( g_Q^{(E,x)} \hat{Q} + g_x \hat{x} \right) \hat{a}^\dagger \hat{a} \enspace ,
\end{equation}
respectively, where in both cases we have generated the coupling between the cavity and the circuit, akin to the optomechanical treatment.

\subsection{Magnetic field sensing}
\label{ss2cm}

For current measurement via magnetic field sensing, we exploit the magnetic fields generated by the current flowing through an inductor and read out the magnetic field using a magnetic-field sensitive parametric cavity, as shown in Fig.~\ref{fig:cavity}b, and exemplified by a microwave transmission line resonator terminated with a DC SQUID~\cite{gerard,gerard2,SQ1,SQog,SQ2}. In this case, we take the parametric cavity to be characterized by a resonance frequency dependent on the magnetic field of a coupled inductor, adding a parallel inductance $L_M$ to the circuits considered thus far, as shown in the right column of Fig.~\ref{fig:cavity}c. We choose this inductance such that $L_M<L$ in order for this additional inductor to act as a relatively low-impedance element for current to flow through.

Noting that the motion of the test masses will alter the current through the inductor $L_M$, thereby changing the magnetic field it generates, we consider the expression of this magnetic field in terms of circuit quantities. For a long solenoid,
\begin{equation}
\label{eq:Bfield}
B = \mu n i^b_{L_M} = \frac{\mu n \Phi^b_{L_M}}{L_M} \enspace , 
\end{equation}
where $\mu$ is the magnetic permeability of the material making up the core of the inductor and $n$ is its turn density. The current flowing through the inductor $i^b_{L_M}$ is expressed in terms of its branch flux $\Phi^b_{L_M}$ in the usual way. We note that expressing the branch flux in terms of circuit degrees of freedom is dependent on the detector configuration, and in the magnetomechanical case, also dependent on the gauge choice. This necessitates some care in expressing $\Phi^b_{L_M}$ in terms of the circuit degrees of freedom.

One can perform an analogous treatment of the current measurement circuits in Fig.~\ref{fig:cavity}c, following the procedures in Appendix~\ref{circuits101}, and confirm that in the magnetomechanical case, the first gauge yields $\Phi^b_{L_M} = -(\Phi + T_v x)$ and the second gauge yields $\Phi^b_{L_M} = -\Phi$, while in the electromechanical case $\Phi^b_{L_M} = \Phi$. The associated Hamiltonians are equivalent to their voltage counterparts except for the addition of an inductive term, $(\Phi^b_{L_M})^2/2L_M$. For the magnetomechanical detector scheme, the Hamiltonians for each gauge are 
\begin{equation}
\label{eq:H1mI}
H_1^{(B,v)} = H_1^{(E,v)} + \frac{(\Phi + T_v x)^2}{2L_M}
\end{equation}
and
\begin{equation}
\label{eq:H2mI}
H_2^{(B,v)} = H_2^{(E,v)} + \frac{\Phi^2}{2L_M} \enspace .
\end{equation}
In the electromechanical case,
\begin{equation}
\label{eq:H1mI}
H^{(B,x)} = H^{(E,x)} + \frac{\Phi^2}{2 L_M} \enspace.
\end{equation}

To include the parametric cavity, we proceed analogously to the voltage measurement case, exchanging $\omega(E)$ for $\omega(B)$ and remaining mindful of the various expressions of the magnetic field appropriate for different detector configurations and gauges. We then find the cavity Hamiltonians appropriate for each gauge in the magnetomechanical detection scheme to be 
\begin{equation}
\label{eq:icav1}
\hat{H}_{\text{cav},1}^{(B,v)} = \hbar \omega_0^B \hat{a}^\dagger \hat{a} - \hbar g^B (\hat{\Phi} + T_v \hat{x}) \hat{a}^\dagger \hat{a} 
\end{equation}
and
\begin{equation}
\label{eq:icav2}
\hat{H}_{\text{cav},2}^{(B,v)} = \hbar \omega_0^B \hat{a}^\dagger \hat{a} - \hbar g^B \hat{\Phi} \hat{a}^\dagger \hat{a} \enspace ,
\end{equation}
where we have defined the coupling constant $g^B = \frac{\eta_B \mu n}{L_M}$, the rescaled resonance frequency $\omega_0^B = \omega(B_0) - \eta_B B_0$, and the cavity's sensitivity to magnetic fields $\eta_B = \frac{d \omega(B)}{dB} \rvert_{B=B_0}$. Similarly, we find the cavity Hamiltonian 
\begin{equation}
\label{eq:icav1}
\hat{H}_{\text{cav}}^{(B,x)} = \hbar \omega_0^{B} \hat{a}^\dagger \hat{a} + \hbar g^B \hat{\Phi} \hat{a}^\dagger \hat{a} 
\end{equation}
for the electromechanical case.

\subsection{Exploiting the optomechanics analogy}
\label{s2om}

To consider the measurement-added noise in these toy models, we incorporate a drive to probe the parametric cavities.
We follow the usual formulation for optomechanical systems~\cite{bowen,clerk,aspel} to include the drive and model the noise associated with quantum fluctuations of the vacuum. Details of this analysis can be found in Appendix \ref{optomechanics}. Altogether, we arrive at the Hamiltonians for voltage and current measurement in both the magnetomechanical and electromechanical detection schemes, accounting for the two gauges in the magnetomechanical case.

For voltage measurement in the magnetomechanical case, we obtain the Hamiltonian
\begin{equation}
\label{eq:HVfinalmm}
\begin{split}
\hat{H'}^{(E,v)}_{i} = -\hbar \Delta \hat{a}^\dagger \hat{a} -& \hbar G_Q^{(E,v)} \hat{Q} \hat{X} \\
&+ \hat{H}_i^{(E,v)} + \hat{H}_B + \hat{H}_{\text{int}} \enspace ,
\end{split}
\end{equation}
where $i = 1,2$ for the two gauge choices. We have defined the detuning $\Delta = \omega_L - \omega_0^{(E,v)}$ with $\omega_L$ the drive frequency, while the Hamiltonians $\hat{H}_B$ and $\hat{H}_{\text{int}}$ describe those of the bath and the bath-cavity coupling, respectively, as defined in Appendix~\ref{optomechanics}. Relevant constants have been collected to define $G_Q^{(E,v)} = \sqrt{2} \alpha g_Q^{(E,v)}$, with $\alpha$ corresponding to the drive strength. Similarly, in the electromechanical case the Hamiltonian is
\begin{equation}
\label{eq:HVfinalem}
\begin{split}
\hat{H'}^{(E,x)} = -\hbar \Delta \hat{a}^\dagger \hat{a} +& \hbar \left( G_Q^{(E,x)} \hat{Q} + G_x \hat{x} \right) \hat{X} \\
&+ \hat{H}^{(E,x)} + \hat{H}_B + \hat{H}_{\text{int}} \enspace ,
\end{split}
\end{equation}
where $G_Q^{(E,x)} = \sqrt{2} \alpha g_Q^{(E,x)}$ and $G_x = \sqrt{2} \alpha g_x$. In both detector schemes, we take the drive strength $\alpha$ to be real, enabling the cavity-circuit coupling to be written in terms of the amplitude quadrature of the cavity, $\hat{X} = (\hat{a} + \hat{a}^\dagger)/\sqrt{2}$. We note there is no loss of generality with this choice of $\alpha$; taking $\alpha$ to be purely imaginary yields a circuit-cavity coupling that goes instead as the phase quadrature of the cavity, $\hat{Y} = \frac{-i}{\sqrt{2}} (\hat{a} - \hat{a}^\dagger)$.

For current measurement in each of the detector configurations and gauges, we find the Hamiltonians 
\begin{equation}
\label{eq:HIfinalmm1}
\begin{split}
\hat{H'}^{(B,v)}_{1} = -\hbar \Delta \hat{a}^\dagger \hat{a} -& \hbar G^B (\hat{\Phi} + T_v \hat{x}) \hat{X} \\
&+ \hat{H}^{(B,v)}_1 + \hat{H}_B + \hat{H}_{\text{int}} \enspace ,
\end{split}
\end{equation}
\begin{equation}
\label{eq:HIfinalmm2}
\begin{split}
\hat{H'}^{(B,v)}_{2} = -\hbar \Delta \hat{a}^\dagger \hat{a} -& \hbar G^B \hat{\Phi} \hat{X} \\
&+ \hat{H}^{(B,v)}_2 + \hat{H}_B + \hat{H}_{\text{int}} \enspace ,
\end{split}
\end{equation}
and
\begin{equation}
\label{eq:HIfinalem}
\begin{split}
\hat{H'}^{(B,x)} = -\hbar \Delta \hat{a}^\dagger \hat{a} +& \hbar G^B \hat{\Phi} \hat{X} \\
&+ \hat{H}^{(B,x)} + \hat{H}_B + \hat{H}_{\text{int}} \enspace ,
\end{split}
\end{equation}
where here we have defined the coupling constant $G^B = \sqrt{2} \alpha g^B$.

In what follows, we use the Hamiltonians in Eqs.~(\ref{eq:HVfinalmm})-(\ref{eq:HIfinalem}) to find and solve the Heisenberg-Langevin equations, enabling the assessment of noise sensitivities.

\section{Combinations of transducers and receivers}
\label{S3}

We now turn our attention to measurement and the consequences of different transducer and receiver combinations, given by the Hamiltonians in Eqs.~(\ref{eq:HVfinalmm})-(\ref{eq:HIfinalem}). For each of these combinations, we solve the Heisenberg equations of motion to find the force noise power spectral density (PSD). We then use the force noise PSD to compare the sensitivities of the different configurations at different frequencies.

\subsection{The equations of motion}

We begin with the usual methods from input-output theory~\cite{colgard}, writing down and solving the Heisenberg equation of motion for the bath modes $\hat{b}(\omega)$.
This enables the equations of motion for the cavity operators~$\hat{a},\hat{a}^\dagger$ to be expressed in terms of the input modes $\hat{b}_{\text{in}}$ and output modes $\hat{b}_{\text{out}}$. The details of this procedure can be found in Appendix~\ref{IOtheory}.

At this stage, it is preferable to recast the equations of motion for the cavity operators in terms of quantities accessible to measurement, namely, the amplitude and phase quadratures of the cavity, $\hat{X}$ and $\hat{Y}$, respectively, where $[\hat{X},\hat{Y}] = i$.
We also define the quadratures of the input and output bath modes as $\hat{X}_{\text{in}} = (\hat{b}_{\text{in}} + \hat{b}_{\text{in}}^\dagger)/\sqrt{2}$, $\hat{Y}_{\text{in}} = -i (\hat{b}_{\text{in}} - \hat{b}_{\text{in}}^\dagger)/\sqrt{2}$, $\hat{X}_{\text{out}} = (\hat{b}_{\text{out}} + \hat{b}_{\text{out}}^\dagger)/\sqrt{2}$, and $\hat{Y}_{\text{out}} = -i (\hat{b}_{\text{out}} - \hat{b}_{\text{out}}^\dagger)/\sqrt{2}$. Using these definitions and the input-output relation given by Eq.~(\ref{eq:inout}) in Appendix~\ref{IOtheory}, it can be verified that each of the input and output quadratures satisfy their own input-output relations of the form
\begin{equation}
\label{eq:XIO}
\hat{X}_{\text{out}} = \hat{X}_{\text{in}} + \sqrt{\kappa}\hat{X}
\end{equation}
and
\begin{equation}
\label{eq:YIO}
\hat{Y}_{\text{out}} = \hat{Y}_{\text{in}} + \sqrt{\kappa}\hat{Y} \enspace ,
\end{equation}
where $\kappa$ corresponds to the cavity decay rate.

By combining the equations of motion for $\hat{a}$ and $\hat{a}^\dagger$ we find the equations of motion for each cavity quadrature. These equations, in combination with the Heisenberg equations of motion for the remaining system operators ($\hat{x}$, $\hat{p}$, $\hat{\Phi}$, and $\hat{Q}$), specify the complete system of equations describing each case. Below, we explicitly show the equations of motion for each detector configuration, measurement scheme, and gauge. We also include an input force $\hat{F}_{\text{in}}$ in the equation for the canonical momentum to account for the impulse we wish to detect. We note that it is appropriate for the input force to act on the canonical momentum rather than the mechanical momentum, as in each instance we consider, the canonical momentum is either strictly the mechanical momentum or a linear combination that includes the mechanical momentum.

We first consider the equations of motion for voltage measurement via electric field sensing. In the magnetomechanical case, the first gauge yields the equations of motion
\begin{equation}
\label{eq:Ev1EOM}
\begin{split}
\frac{d\hat{x}}{dt} &= \frac{\hat{p}}{m} - \frac{T_v}{m} \hat{Q} \enspace , \\
\frac{d\hat{p}}{dt} &= -k \hat{x} + \hat{F}_{\text{in}} \enspace , \\
\frac{d\hat{\Phi}}{dt} &= -\hbar G_Q^{(E,v)} \hat{X} - \frac{T_v}{m} \hat{p} + \left( \frac{1}{C_L} + \frac{T_v^2}{m} \right) \hat{Q} \enspace , \\
\frac{d\hat{Q}}{dt} &= -\frac{\hat{\Phi}}{L} \enspace , \\\frac{d\hat{X}}{dt} &= -\Delta \hat{Y} - \frac{\kappa}{2} \hat{X} - \sqrt{\kappa} \hat{X}_{\text{in}} \enspace , \\
\frac{d\hat{Y}}{dt} &= \Delta \hat{X} + G_Q^{(E,v)} \hat{Q} - \frac{\kappa}{2} \hat{Y} - \sqrt{\kappa} \hat{Y}_{\text{in}} \enspace .
\end{split}
\end{equation}
It is convenient to rewrite these equations in a more compact form. Defining the matrix
\begin{equation}
\label{eq:MEv1}
    \underline{\underline{M_1}}^{(E,v)} = \begin{bmatrix}
    0& \frac{1}{m} & 0 & \textrm{-}\frac{T_v}{m} & 0 & 0\\[0.6em]
    \textrm{-}k& 0 & 0 &0 & 0 & 0\\[0.6em]
    0& \textrm{-}\frac{T_v}{m} &  0   &\frac{1}{C^\prime_L} & \textrm{-}\hbar G_Q^{(E,v)} & 0\\[0.6em]
    0& 0 & \textrm{-} \frac{1}{L} & 0  & 0 & 0\\[0.6em]
    0& 0 & 0 &0 & \textrm{-}\frac{\kappa}{2} & \textrm{-}\Delta\\[0.6em]
    0& 0 & 0 & G_Q^{(E,v)} & \Delta & \textrm{-}\frac{\kappa}{2}
    \end{bmatrix} \enspace ,
\end{equation}
we can rewrite Eq.~(\ref{eq:Ev1EOM}) as
\begin{equation}
\label{eq:EOMex}
\frac{d}{dt} \underline{\hat{Z}} = \underline{\underline{M_1}}^{(E,v)} \underline{\hat{Z}} + \underline{\hat{Z}_{\text{in}}} \enspace ,
\end{equation}
where we have defined the vector of operators $\underline{\hat{Z}}~=~\lbrace \hat{x},\hat{p},\hat{\Phi},\hat{Q},\hat{X},\hat{Y} \rbrace$ and the vector of inputs $\underline{\hat{Z}_{\text{in}}}~=~\lbrace 0,\hat{F}_{\text{in}},0,0,\textrm{-}\sqrt{\kappa} \hat{X}_{\text{in}},\textrm{-}\sqrt{\kappa} \hat{Y}_{\text{in}}\rbrace$. We also define the capacitance $C_L^\prime$ for convenience to be $\frac{1}{C_L^\prime} =  \frac{1}{C_L}+\frac{T_v^2}{m}$. The equations of motion for the second gauge can be similarly represented via Eq.~(\ref{eq:EOMex}) with the matrix $\underline{\underline{M_2}}^{(E,v)}$, defined~as
\begin{equation}
\label{eq:MEv2}
    \underline{\underline{M_2}}^{(E,v)} = \begin{bmatrix}
    0& \frac{1}{m} & 0 &0 & 0 & 0\\[0.6em]
    \textrm{-}k^\prime& 0 & \frac{T_v}{L} & 0 & 0 & 0\\[0.6em]
    0& 0 & 0 & \frac{1}{C_L}  & \textrm{-}\hbar G_Q^{(E,v)} & 0\\[0.6em]
    \frac{T_v}{L}& 0 & \textrm{-} \frac{1}{L} &0 & 0 & 0\\[0.6em]
    0& 0 & 0 &0 & \textrm{-} \frac{\kappa}{2} & \textrm{-} \Delta\\[0.6em]
    0& 0 & 0 & G_Q^{(E,v)} & \Delta & \textrm{-}\frac{\kappa}{2} \\[0.6em]
    \end{bmatrix} \enspace ,
\end{equation}
where we have defined for convenience $k^\prime = k + \frac{T_v^2}{L}$. Likewise, the electromechanical case is described by the matrix
\begin{equation}
\label{eq:MEx}
    \underline{\underline{M}}^{(E,x)}=
  \begin{bmatrix}
    0& \frac{1}{m} & 0 &0 & 0 & 0\\[0.6em]
    \textrm{-}k_{\rm eff}& 0 & 0 & \frac{T_x}{C_P} & \textrm{-}\hbar G_x & 0\\[0.6em]
    \textrm{-}\frac{T_x}{C_P} & 0 & 0 & \frac{1}{C_{\rm eff}}  & \hbar G_Q^{(E,x)} & 0\\[0.6em]
    0 & 0 & \textrm{-} \frac{1}{L} &0 & 0 & 0\\[0.6em]
    0& 0 & 0 &0 & \textrm{-} \frac{\kappa}{2} & \textrm{-} \Delta\\[0.6em]
    \textrm{-}G_x & 0 & 0 & \textrm{-}G_Q^{(E,x)} & \Delta & \textrm{-}\frac{\kappa}{2} \\[0.6em]
    \end{bmatrix} \enspace .
\end{equation}

We can analogously describe the equations of motion for the current measurement scheme. In the magnetomechanical case, we find for the two gauges
\begin{equation}
\begin{split}
\label{eq:MBv1}
    \underline{\underline{M_1}}^{(B,v)} =\begin{bmatrix}
    0& \frac{1}{m} &0 &\textrm{-}\frac{T_v}{m} & 0 & 0\\[0.6em]
    \textrm{-}k^\prime_M& 0 & \textrm{-}\frac{T_v}{L_M} & 0 & \hbar G^B T_v & 0\\[0.6em]
    0& \textrm{-}\frac{T_v}{m} &0 & \frac{1}{C_L^\prime} & 0 & 0\\[0.6em]
     \textrm{-}\frac{T_v}{L_M} & 0 & \textrm{-} \frac{1}{L^\prime} &0 & \hbar G^B & 0\\[0.6em]
    0& 0 & 0 &0 & \textrm{-}\frac{\kappa}{2} & \textrm{-}\Delta\\[0.6em]
    G^B T_v& 0 & G^B &0 & \Delta & \textrm{-}\frac{\kappa}{2} \\[0.6em]
    \end{bmatrix} \enspace ,
    \end{split}
\end{equation}
and
\begin{equation}
\label{eq:MBv2}
    \underline{\underline{M_2}}^{(B,v)} = \begin{bmatrix}
    0& \frac{1}{m} & 0 &0 & 0 & 0\\[0.6em]
     \textrm{-}k^\prime& 0 & \frac{T_v}{L} & 0 &0 & 0\\[0.6em]
    0& 0 & 0 &\frac{1}{C_L} & 0 & 0\\[0.6em]
     \frac{T_v}{L}& 0 &  \textrm{-} \frac{1}{L^\prime} &0 & \hbar G^B & 0\\[0.6em]
    0& 0 & 0 &0 & \textrm{-}\frac{\kappa}{2} & \textrm{-}\Delta\\[0.6em]
    0& 0 & G^B &0 & \Delta & \textrm{-}\frac{\kappa}{2} \\[0.6em]
    \end{bmatrix} \enspace ,
\end{equation}
where we have defined for convenience the quantities $k^\prime_M = k + \frac{T_v^2}{L_M}$ and $ \frac{1}{L^\prime}= \frac{1}{L}+ \frac{1}{L_M} $. For the electromechanical detector configuration, we have 
\begin{equation}
\label{eq:MBx}
    \underline{\underline{M}}^{(B,x)}=
  \begin{bmatrix}
    0& \frac{1}{m} & 0 &0 & 0 & 0\\[0.6em]
    \textrm{-}k_{\rm eff}& 0 & 0 & \frac{T_x}{C_P} & 0 & 0\\[0.6em]
    \textrm{-}\frac{T_x}{C_P} & 0 & 0 & \frac{1}{C_{\rm eff}}  & 0 & 0\\[0.6em]
    0 & 0 & \textrm{-} \frac{1}{L^\prime} &0 & \textrm{-}\hbar G^B & 0\\[0.6em]
    0& 0 & 0 &0 & \textrm{-} \frac{\kappa}{2} & \textrm{-} \Delta\\[0.6em]
    0& 0 & \textrm{-}G^B & 0 & \Delta & \textrm{-}\frac{\kappa}{2} \\[0.6em]
    \end{bmatrix} \enspace .
\end{equation}

\subsection{Assembling the noise PSD}

These linear equations are straightforward to solve in the frequency domain. Defining the Fourier transform of our relevant operators as follows
\begin{equation}
\begin{split}
\hat{f}(\nu) &= \frac{1}{\sqrt{2 \pi}} \int_{-\infty}^{\infty} \hat{f}(t) e^{-i\nu t} dt \\ \hat{f}(t) &= \frac{1}{\sqrt{2 \pi}} \int_{-\infty}^\infty \hat{f}(\nu) e^{i\nu t} d\nu
\end{split} \enspace ,
\end{equation}
where the frequency dependence of $\hat{f}(\nu)$ is explicit, the time derivatives in the equation of motion given by Eq.~(\ref{eq:EOMex}) for each matrix $\underline{\underline{M}}$ simply transform as $\frac{d\hat{f}(t)}{dt} \rightarrow i \nu \hat{f}(\nu)$.

Solving the systems of equations given by Eq.~(\ref{eq:EOMex}) in the frequency domain, for zero detuning ($\Delta = 0$), yields a general solution of the form
\begin{equation}
\label{eq:GS}
\underline{\hat{Z}}(\nu) = \left( i \nu \underline{\underline{\mathbb{1}}} - \underline{\underline{M}} \right)^{-1} \underline{\hat{Z}_{\text{in}}}(\nu) \enspace ,
\end{equation}
where we focus on the solutions for the phase quadrature of the cavity $\hat{Y}(\nu)$, given by the final row of Eq.~(\ref{eq:GS}). The circuit degrees of freedom are coupled to $\hat{X}$, hence, the information about the circuit, the mechanics, and ultimately $\hat{F}_\text{in}$ are found in the equation of motion and solution of its conjugate, $\hat{Y}$ (as opposed to $\hat{X}$, the other quantity accessible to measurement).

In the magnetomechanical system, the solutions found in each gauge are identical, sans the solutions for $\hat{p}$ and $\hat{\Phi}$. This is unsurprising, since these quantities are those affected by the unitary [Eq.~(\ref{eq:GTmech})] that connects the gauges. This is true in both the voltage and current measurement cases.
For both the magnetomechanical and electromechanical detector schemes, the solutions for $\hat{Y}(\nu)$ may then be used in the input-output relation, Eq.~(\ref{eq:YIO}), to find the output quadrature $\hat{Y}_{\text{out}}(\nu)$. These solutions are explicitly shown in Appendix~\ref{noisePSD}.

We then use the solution for the output quadrature $\hat{Y}_{\text{out}}(\nu)$ to assess the noise sensitivity. We define the force estimator as the output phase quadrature in force units (i.e., we divide $\hat{Y}_{\text{out}}$ by the coefficient of $\hat{F}_{\text{in}}$):
\begin{equation}
\hat{F}_E(\nu) = \hat{F}_{\text{in}} + \beta(\nu) \hat{X}_{\text{in}} + \gamma(\nu) \hat{Y}_{\text{in}} \enspace ,
\end{equation}
where $\beta$ represents the coefficient for the measurement backaction noise term  and $\gamma$ corresponds to the shot noise term in this measurement, as we are measuring the $\hat{Y}$ quadrature.

The frequency dependence of the noise sensitivity is given by the force noise power spectral density (PSD):
\begin{equation}
\label{eq:SFFdef}
S_{\text{FF}}(\nu) = \int_{-\infty}^{\infty} \braket{\hat{F}_E^\dagger(\nu)  \hat{F}_E(\nu')} d\nu' \enspace .
\end{equation}
To evaluate the noise PSD in Eq.~(\ref{eq:SFFdef}), we note the following regarding the resulting noise correlation functions:
\begin{equation}
\begin{split}
	\braket{\hat{X}_{\text{in}}^\dagger(t)  \hat{X}_{\text{in}}(t')} &= \braket{\hat{Y}_{\text{in}}^\dagger(t)  \hat{Y}_{\text{in}}(t')} = \frac{1}{2} \delta (t-t') \\
 \braket{\hat{F}_{\text{in}}^\dagger(t)  \hat{F}_{\text{in}}(t')} & = N_{\rm BM} \delta (t-t')
\end{split} \enspace ,
\end{equation}
where we have taken the vacuum fluctuations of the cavity to be white noise and the input signal noise from the mechanics to be thermal noise. We consider here an Ohmic model for thermal noise corresponding to Brownian motion with noise amplitude $N_{\rm BM}$. We assume there is no correlation between the signal noise and vacuum fluctuations, and no correlation between the vacuum fluctuations.
Taking all of the above into consideration, Eq.~(\ref{eq:SFFdef}) can be rewritten as
\begin{equation}
\label{eq:SFF}
S_{\text{FF}}(\nu) = \frac{|\beta(\nu)|^2}{2} + \frac{|\gamma(\nu)|^2}{2} + N_{\text{BM}} \enspace .
\end{equation}

In what follows, we list the noise PSD expressions for each combination of detector configuration and measurement scheme (recalling that the solutions across gauges are identical) in terms of relevant susceptibilities. In each case, we optimize the coupling strength (and as a result, the drive strength) to balance the backaction and shot noise terms at some fixed frequency, thereby identifying the SQL at this frequency for each case. Then, we compare the performances of different detector and measurement combinations and discuss the features of these noise PSD expressions.

It is convenient to first define the frequencies
\begin{equation}
\label{eq:scales}
\begin{split}
\omega_m^2 = \frac{k}{m},\hspace{1mm} &\omega_{c (\rm ce)}^2 = \frac{1}{L C_{L ({\rm eff})}},\hspace{1mm} \omega_{l ({\rm le})}^2 = \frac{1}{L_M C_{L ({\rm eff})}} \\
&\delta_v^2 = \frac{T_v^2}{m L}, \hspace{1mm} \delta_x^2 = \frac{C_{\rm eff} T_x^2}{m C_P^2} \enspace ,
\end{split}
\end{equation}
as well as the cavity and bare mechanical susceptibilities $\chi_{\kappa}$ and  $\chi_m$, respectively:
\begin{equation}
\label{eq:commonchi}
\chi_\kappa = \frac{-\sqrt{\kappa}}{\frac{\kappa}{2} + i \nu},\hspace{1mm} \chi_m = \frac{-1}{\nu^2 - \omega_m^2} \enspace .
\end{equation}
For the magnetomechanical detector scheme, voltage measurement via electric field sensing yields the noise PSD
\begin{equation}
\label{magnetic-voltage}
\begin{split}
S_{\text{FF}}^{(E,v)} = &\frac{\hbar^2 {G_Q^{(E,v)}}^2 m |\chi_{\kappa}|^2 }{2 L \delta_v^2 \chi_m^2 \nu^2} \\
&+ \frac{m L}{ 2 {G_Q^{(E,v)}}^2 \delta_v^2 |\chi_{\kappa}|^2 \chi_m^2 {\chi_{\rm lc}^{(E,v)}}^2 \nu^2} +   N_{\rm BM} \enspace ,
\end{split}
\end{equation}
where we define the circuit susceptibility as
\begin{equation}
\label{eq:chimmv}
\chi_{\rm lc}^{(E,v)} = \frac{-1}{\nu^2 - \omega_c^2 - \delta_v^2 (1-\chi_m \omega_m^2)} \enspace .
\end{equation}
Upon balancing the backaction and shot noise terms for some fixed frequency, we find the optimized coupling strength
\begin{equation}
\label{eq:GQvSQL}
{G_Q^{(E,v)}}^2 = \frac{L}{\hbar |\chi_\kappa|^2 |\chi_{\rm lc}^{(E,v)}|} \enspace .
\end{equation}
For current measurement via magnetic field sensing, we find the noise PSD
\begin{equation}
\label{magnetic-current}
\begin{split}
S_{\text{FF}}^{(B,v)} = &\frac{\hbar^2 {G^B}^2 m L |\chi_\kappa|^2}{2 \delta_v^2 {\chi_m^\prime}^2} \\
&+ \frac{m}{2 {G^B}^2 L \delta_v^2 |\chi_\kappa|^2 {\chi_m^\prime}^2 {\chi_{\rm lc}^{(B,v)}}^2 \omega_c^4} +    N_{\rm BM} \enspace ,
\end{split}
\end{equation}
where we have instead defined the dressed mechanical susceptibility
\begin{equation}
\chi_m^\prime = \frac{-1}{\nu^2 - \omega_m^2 - \delta_v^2}
\end{equation}
as well as the circuit susceptibility in this case as
\begin{equation}
\label{eq:chimmc}
\chi_{\rm lc}^{(B,v)} = \frac{-1}{\nu^2 - \omega_c^2 - \omega_l^2 + \delta_v^2 \omega_c^2 \chi_m^\prime} \enspace .
\end{equation}
In this instance, we find the optimized coupling strength to be 
\begin{equation}
\label{eq:GBvSQL}
{G^B}^2 = \frac{1}{\hbar |\chi_\kappa|^2 |\chi_{\rm lc}^{(B,v)}| L \omega_c^2} \enspace .
\end{equation}
We note that in the magnetomechanical detector scheme, the various mechanical and circuit susceptibilities are gauge-dependent --- just as the canonical momenta are different across the two gauges, so are the associated response functions. The physical meaning of these susceptibilities therefore depends on the gauge choice. We emphasize that the noise PSDs are gauge-independent, however, expressing the noise PSDs in terms of these gauge-dependent functions yields different functional forms of the noise PSDs. We direct the reader to Appendix~\ref{Full PSD} for the explicit forms of the noise PSDs written in terms of the relevant frequencies defined in Eq.~(\ref{eq:scales}).

For the electromechanical detector scheme, voltage measurement via electric field sensing yields the noise PSD
\begin{equation}
\centering
\label{electric-voltage}
\begin{split}
S_{\text{FF}}^{(E,x)} &\approx \frac{\hbar^2 {G_Q^{(E,x)}}^2 m |\chi_\kappa|^2}{2 \delta_x^2 \chi_m^2 L \omega_{\rm ce}^2} \\
&+ \frac{ m L }{ 2 {G_Q^{(E,x)}}^2 \delta_x^2 |\chi_\kappa|^2 \chi_m^2 {\chi_{\rm lc}^{(E,x)}}^2 \omega_{\rm ce}^2} + N_{\rm BM} \enspace , 
\end{split}
\end{equation} 
where in this case we define the circuit susceptibility
\begin{equation}
\label{eq:chiemv}
\chi_{\rm lc}^{(E,x)} = \frac{-1}{\nu^2 - \omega_{\rm ce}^2 + \delta_x^2 \omega_{\rm ce}^2 \chi_m} \enspace. 
\end{equation}
In Eq.~(\ref{electric-voltage}), we have taken $G_x \rightarrow 0$, as for our chosen parameters, $G_Q^{(E,x)} \gg G_x$. The exact noise PSD, including the contributions from $G_x$, can be found in Appendix~\ref{Full PSD}. We note that this contribution is included in the numerics we present in Sections~\ref{comparison} and \ref{details}, which confirms the negligible contribution from $G_x$. Therefore, we optimize the coupling to balance the backaction and shot noise terms with respect to $G_Q^{(E,x)}$ only, using Eq.~(\ref{electric-voltage}). We find this coupling strength to be
\begin{equation}
\label{eq:GQxSQL}
{G_Q^{(E,x)}}^2 = \frac{L}{\hbar |\chi_\kappa|^2 |\chi_{\rm lc}^{(E,x)}|} \enspace .
\end{equation}
For current measurement via magnetic field sensing we find the expression
\begin{equation}
\centering
\label{electric-current}
\begin{split}
S_{\text{FF}}^{(B,x)} &= \frac{\hbar^2 {G^B}^2 m L \omega_{\rm ce}^2 |\chi_\kappa|^2 \left( 1 - \delta_x^2 \chi_m \right)^2}{2 \delta_x^2 \chi_m^2 \nu^2} \\
&+ \frac{ m }{ 2 {G^B}^2 L \delta_x^2 |\chi_\kappa|^2 \chi_m^2 {\chi_{\rm lc}^{(B,x)}}^2 \omega_{\rm ce}^2 \nu^2} +   N_{\rm BM} \enspace ,
\end{split}
\end{equation}
where we define the circuit susceptibility 
\begin{equation}
\label{eq:chiemc}
\chi_{\rm lc}^{(B,x)} = \frac{-1}{\nu^2 - \omega_{\rm ce}^2 - \omega_{\rm le}^2 + \delta_x^2 (\omega_{\rm ce}^2 + \omega_{\rm le}^2) \chi_m} \enspace.
\end{equation}
To balance backaction noise and shot noise, we find the optimized coupling
\begin{equation}
\label{eq:GBxSQL}
{G^B}^2 = \frac{1}{\hbar |\chi_\kappa|^2 |\chi_{\rm lc}^{(B,x)}|  |1 - \delta_x^2 \chi_m| L \omega_{\rm ce}^2} \enspace .
\end{equation}

In what follows, we examine the noise PSDs given by Eqs.~(\ref{magnetic-voltage}), (\ref{magnetic-current}), (\ref{electric-voltage}), and (\ref{electric-current}), comparing their behavior to the standard optomechanical case and examining their specific functional features.

\subsection{Comparison to optomechanical systems}
\label{comparison}

Our goal is to look for signals which have broad characteristics in frequency space (i.e., an impulse in the time domain), requiring an integration of the noise over a frequency band to be able to predict a signal to noise ratio~(SNR). Thus, our interest is in the broadband sensitivity of the noise PSDs. Additionally, if we can directly access a QND-like variable, such as the velocity of the mechanical system, we expect the measurement backaction to decrease over a broad frequency spectrum. Therefore, we restrict our discussion to the broadband frequency response of the noise PSDs in an effort to understand the best readout strategies for certain kinds of transducers subject to broadband signals.

\begin{figure}
\begin{center}
\includegraphics[width=0.48 \textwidth]{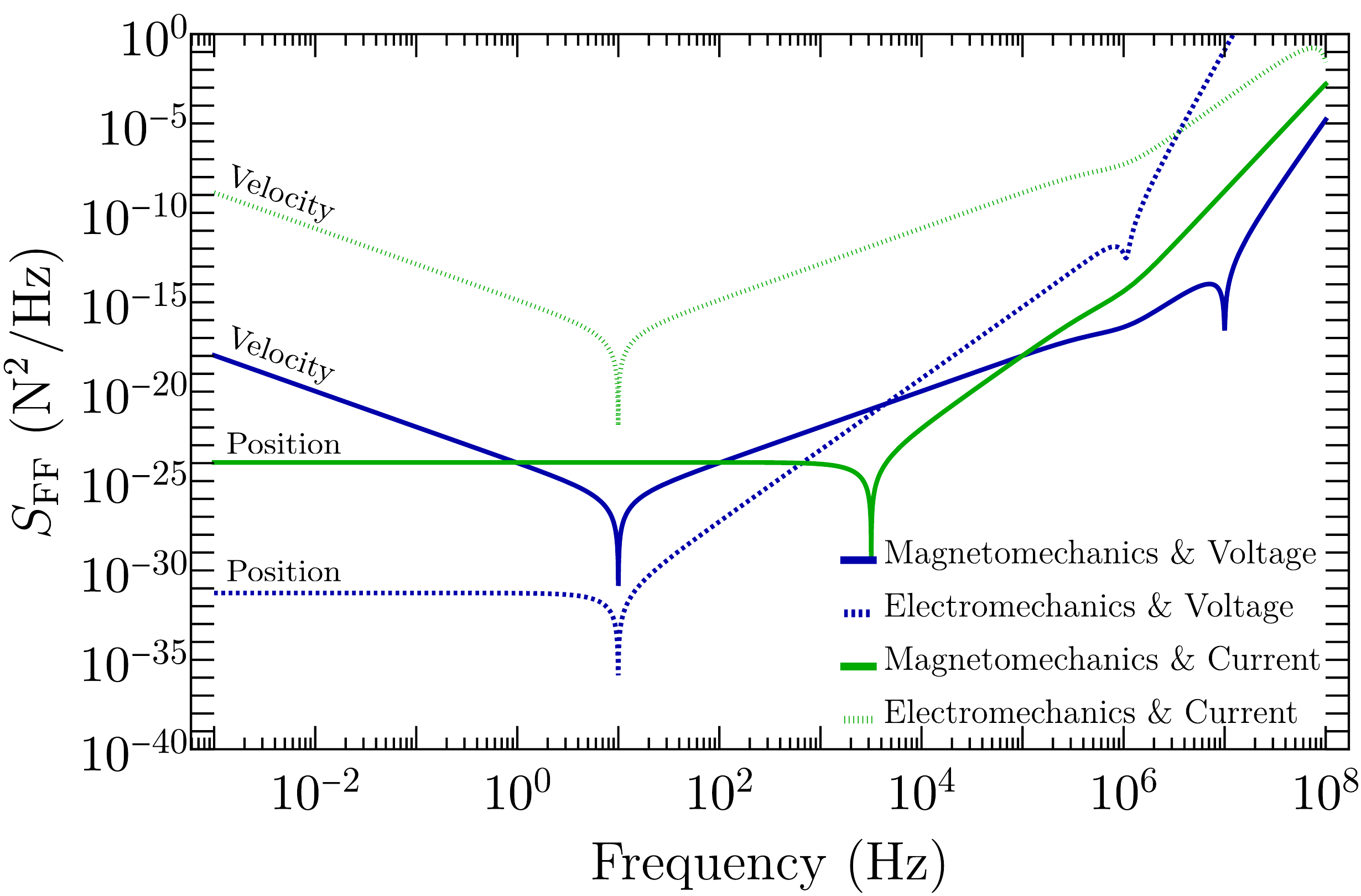}
\caption{Total measurement-added noise for each detector configuration and readout choice. The dark blue curves correspond to voltage readout via electric field sensing for the magnetomechanical and electromechanical configurations. Here, the magnetomechanical system displays a backaction-evading characteristic absent in the electromechanical case. The green curves correspond to current readout via magnetic field sensing for each detector configuration. However, in this case the electromechanical system displays a backaction-evading feature rather than the magnetomechanical case. The parameters used for generating these plots are as follows: detector mass $m = 1 ~ {\rm g}$, mechanical resonance frequency $\omega_m/2\pi = 10 ~ {\rm Hz} $, cavity decay rate $\kappa/2\pi = 1 ~ {\rm MHz}$, inductance  $L = 10 ~ {\rm \mu H}$, mutual inductance $L_M = 1 ~ {\rm n H}$, circuit resonance frequencies $\omega_c/2\pi = 10$ MHz and $\omega_{\rm ce}/2\pi \approx 1$ MHz, capacitance $C_P = 25  ~ {\rm fF}$, with transducer constants $T_v = 2 ~ {\rm T \cdot m}$ and $T_x = \textrm{-}10^{-10} ~ {\rm C/m}$. The coupling coefficients are fixed to their SQL values for a target frequency of $1$ MHz, as given by Eqs.~(\ref{eq:GQvSQL}), (\ref{eq:GBvSQL}), (\ref{eq:GQxSQL}), and (\ref{eq:GBxSQL}), with  $G_x \approx T_x G_Q^{(E,x)}$.}
\label{fig:comparison}
\end{center}
\end{figure}

\begin{figure}[b]
\begin{center}
\includegraphics[width=0.48 \textwidth]{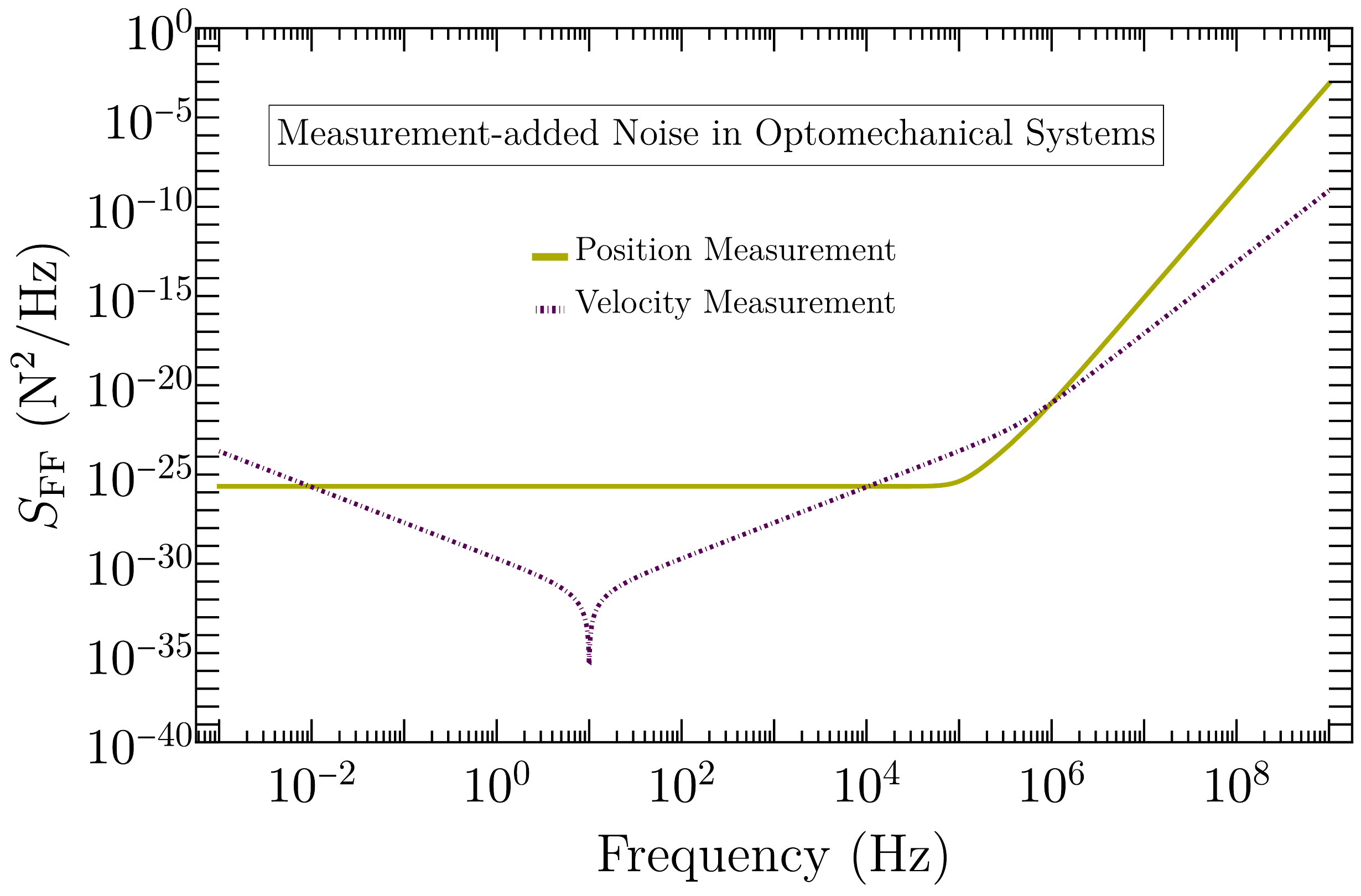}
\caption{The noise PSD representing total measurement-added noise in an optomechanical system is plotted for both position- and velocity-sensing protocols while operating at the optimal power for position sensing with a $0.1 \text{ MHz}$ target frequency (using Eq.~\ref{Optimized G}). 
This is derivative of work from Refs.~\cite{sohitri, ghosh2022combining}. The noise for velocity sensing is lower than position sensing across a broad frequency range, with the functional dependence $\nu^{-2}$ below resonance and $\nu^2$ above. The optomechanical coupling strengths in these techniques are related by the velocity coupling coefficient $G^\prime \rightarrow G/( m \kappa)$, with position coupling coefficient $G/2\pi \approx 10^{23} \text{  Hz/m}$, mechanical frequency $\omega_m/2\pi = 10 \text{  Hz}$, cavity decay rate $\kappa/2\pi = 1 \text{ MHz}$, detector mass $m = 1  \text{ g}$, and mechanical damping rate $\mu/2\pi = 0.1 \text{ mHz}$.}
\label{previous}
\end{center}
\end{figure}

In Fig.~\ref{fig:comparison}, we show the total measurement-added noise for each detector circuit and measurement readout combination, given by Eqs.~(\ref{magnetic-voltage}), (\ref{magnetic-current}), (\ref{electric-voltage}), and (\ref{electric-current}).
We note that we have taken the thermal noise $N_{\rm BM}$ affecting the mechanical oscillator to be negligible. In addition, we have fixed the coupling coefficients to their optimized SQL values, given by Eqs.~(\ref{eq:GQvSQL}), (\ref{eq:GBvSQL}), (\ref{eq:GQxSQL}), and (\ref{eq:GBxSQL}), for a target frequency of $1$ MHz. 
Taken together, the total noise and its associated behavior are particularly relevant in the context of the standard optomechanical position-sensing problem as well as previous work on velocity sensing in optomechanical systems~\cite{sohitri, ghosh2022combining}. For convenience of comparison, we include Fig.~\ref{previous}, which shows the expected noise PSD for both position- and velocity-sensing scenarios in an optomechanical analogue, specifically for use with a broadband signal. Details associated with this plot can be found in Appendix~\ref{Previous-Work}.

Upon comparison, it is immediately clear that current readout of the magnetomechanical detector scheme and voltage readout of the electromechanical detector scheme share similarities with the noise PSD for standard optomechanical position sensing, specifically the `flat at low frequency' feature. Furthermore, voltage readout of the magnetomechanical detector scheme and current readout of the electromechanical detector scheme bear striking similarity to the noise PSD for velocity sensing. Both show a decrease in total noise near the mechanical resonance, and share the same frequency dependence in this region, namely, going as $\nu^{-2}$ below resonance and $\nu^2$ above resonance.

We understand these similarities by using the fundamental relations describing how the mechanics are tranduced to an electrical signal in each detector configuration, outlined in Section~\ref{S1}. In the magnetomechanical case, the flux (comparable to current) is proportional to the position $x$ of the mechanical oscillator, while in the electromechanical case, charge (comparable to voltage) is proportional to position. Thus, by coupling the parametric cavity to a specific circuit degree of freedom, both of these readout schemes access the position of the mechanical oscillator. Alternatively, it is voltage in the magnetomechanical case and current in the electromechanical case which are directly proportional to velocity. Therefore, electromechanical current readout and magnetomechanical voltage readout directly access the velocity of the oscillator, providing a way to reduce the measurement-added backaction noise over certain bandwidths of frequencies.

These results indicate that if we want to attain a QND-like measurement using an electrical circuit setup in the microwave domain, we need to combine a magnetomechanical or electromechanical detection scheme with the appropriate measurement readout. Namely, voltage readout for a magnetomechanical detector and current readout for an electromechanical detector may yield a QND-like measurement with reduced backaction over a broad frequency range. We note that upon comparing the relative scale of the noise between these cases, voltage readout of the magnetomechanical detector scheme yields a considerably lower noise floor than current readout of the electromechanical detector scheme. However, the total noise and its associated resonances (discussed in Section~\ref{details}) are dependent on the system parameters and relevant frequencies.

\subsection{Details of the noise curves}
\label{details}

\begin{figure*}
\begin{center}
\includegraphics[scale=0.46]{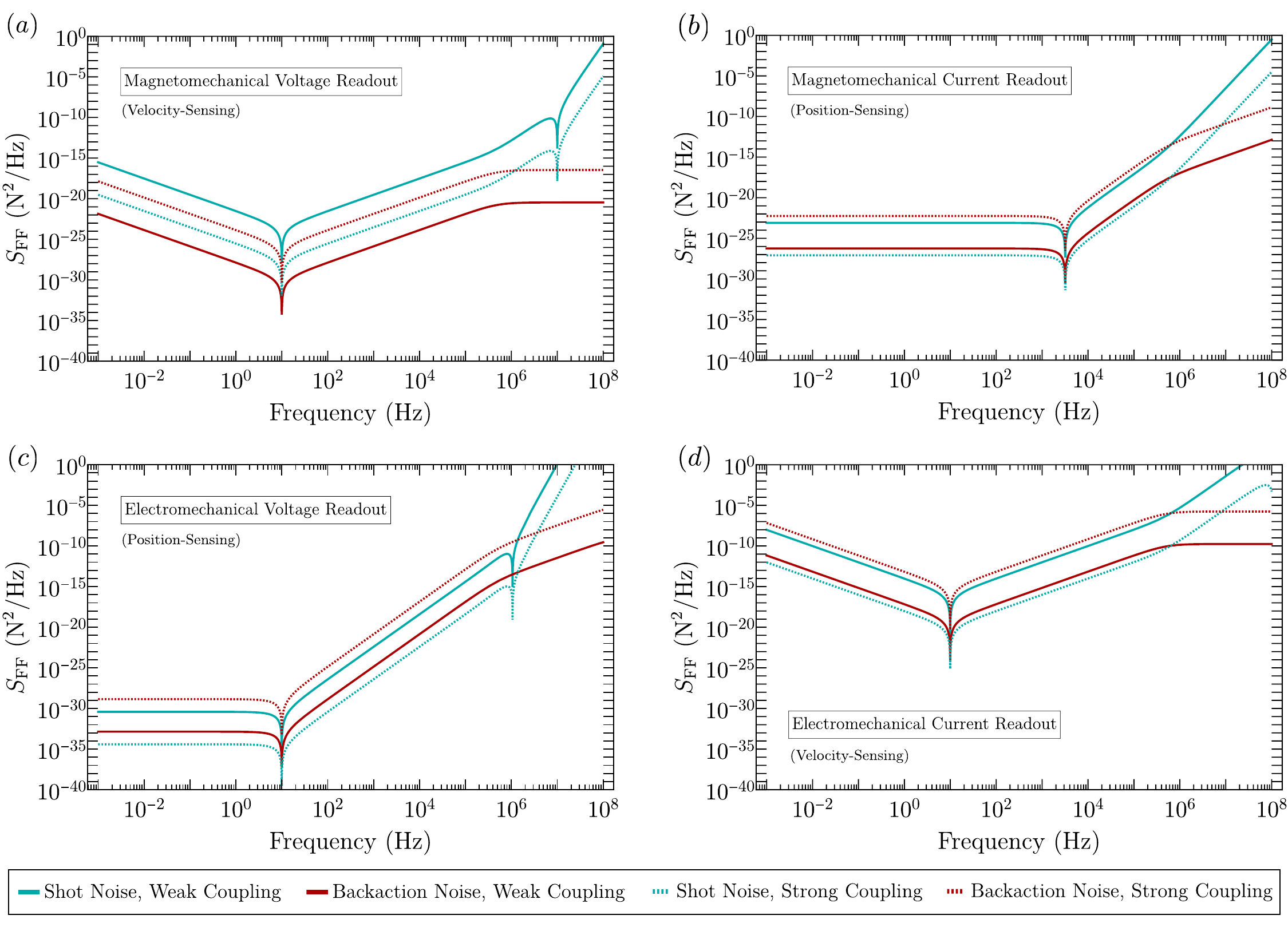}
\caption{Shot noise (light blue) and backaction noise (dark red) curves at different coupling strengths for the magnetomechanical and electromechanical setups. In all instances solid curves correspond to weaker coupling strengths relative to the dashed curves, as indicated in the legend. The weak coupling strengths correspond to the values $G_Q^{(E,v)}/2\pi = G_Q^{(E,x)}/2\pi = 10^{23} \text{ Hz/C}$ and $G^B/2\pi = 10^{23} \text{ Hz/Wb}$, while strong coupling strengths correspond to the values $G_Q^{(E,v)}/2\pi = G_Q^{(E,x)}/2\pi = 10^{25} \text{ Hz/C}$ and $G^B/2\pi = 10^{25} \text{ Hz/Wb}$. Plots in (a) and (b) show the curves for the magnetomechanical detector configuration, where (a) represents the electric field-dependent parametric cavity or voltage readout and (b) represents the magnetic field-dependent parametric cavity or current readout. Plots in (c) and (d) show the curves for the electromechanical detector configuration, where (c) represents the electric field-dependent parametric cavity or voltage readout and (d) represents the magnetic field-dependent parametric cavity or current readout. 
The parameters used for generating these plots are as follows: detector mass $m = 1 ~ {\rm g}$, mechanical resonance frequency $\omega_m/2\pi = 10 ~ {\rm Hz} $, cavity decay rate $\kappa/2\pi = 1 ~ {\rm MHz}$, inductance $L = 10 ~ {\rm \mu H}$, mutual inductance $L_M = 1 ~ {\rm n H}$, circuit resonance frequencies $\omega_c/2\pi = 10$ MHz and $\omega_{\rm ce}/2\pi \approx 1$ MHz, capacitance $C_P = 25  ~ {\rm fF}$, with transducer constants $T_v = 2 ~ {\rm T \cdot m}$ and $T_x = \textrm{-}10^{-10} ~ {\rm C/m}$ and coupling coefficient $G_x \approx T_x G_Q^{(E,x)}$.}
\label{fig:noise}
\end{center}
\end{figure*}

Upon closer inspection of Eqs.~(\ref{magnetic-voltage}), (\ref{magnetic-current}), (\ref{electric-voltage}), and (\ref{electric-current}), we find that each of the noise PSDs has a term inversely proportional to the coupling coefficient squared, i.e., ${G_Q^{(E,x)}}^2$ and ${G_Q^{(E,v)}}^2$ for the voltage measurement scenarios or ${G^B}^2$ for the current measurement scenarios. These terms originate from the $\braket{\hat{Y}_{\rm in}^2}$ contribution to the noise PSD, corresponding to the factor $\frac{|\gamma|^2}{2}$ in Eq.~(\ref{eq:SFF}). As we are interested in monitoring the $\hat{Y}_{\rm out}$ quadrature, we understand this contribution as shot noise --- it constitutes the statistical counting error at the output port. In addition, each noise PSD has another term which is directly proportional to the coupling coefficient squared. These terms arise from the $\braket{\hat{X}_{\rm in}^2}$ contribution to the noise PSD, corresponding to the factor $\frac{|\beta|^2}{2}$ in Eq.~(\ref{eq:SFF}). These terms form the basis of backaction noise on the measurement of the output phase quadrature. Fig.~\ref{fig:noise} shows curves representing these shot noise and backaction noise contributions to the total noise PSD, for each of the different detector and readout combinations. Also included in Fig.~\ref{fig:noise} is a comparison of two different coupling strengths for each noise contribution, where we see that the shot noise decreases with an increase in coupling strength (i.e., a stronger drive) while the backaction noise increases.

We also note the presence of various resonances.
For voltage readout of the magnetomechanical system, shown in Fig.~\ref{fig:noise}a and described by Eq.~(\ref{magnetic-voltage}), the backaction noise term exhibits a resonance at the mechanical frequency~$\omega_m$ while the shot noise term has two resonances: one near $\omega_m$ and the other near the self-resonance of the detector circuit $\omega_c$. This is a consequence of our chosen parameters which result in $\omega_l^2 \gg \omega_c^2 \gg \delta_v^2 \gg \omega_m^2$, where the frequency $\delta_v$ represents the shift in the mechanical resonance due to the circuit coupling. Similarly, for current readout in the magnetomechanical system, shown in Fig.~\ref{fig:noise}b and described by Eq.~(\ref{magnetic-current}), resonances occur near~$\delta_v$ in both the backaction and shot noise terms. Here the mechanics are dressed by the circuit, yielding an effective mechanical resonance at $\delta_v$ with a negligible contribution from $\omega_m$.

For the electromechanical system, described by Eqs.~(\ref{electric-voltage}) and (\ref{electric-current}) and shown in Figs.~\ref{fig:noise}c and \ref{fig:noise}d, both voltage and current readout demonstrate resonances near~$\omega_m$. In addition, a resonance near the circuit's self-resonance $\omega_{\text{ce}}$ is visible in the voltage measurement case in Fig.~\ref{fig:noise}c. As before, the locations of these resonances are a consequence of our chosen parameters and the coupling between the mechanical system and the circuit, where we note $G_Q^{(E,x)} \gg G_x$ and $\omega_{\text{le}}^2 \gg \omega_{\text{ce}}^2 \gg \omega^2_m \gg \delta_x^2$. Here again the mechanics dress the circuit; however, in contrast to the magnetomechanical case, the contribution from $\delta_x$ is negligible in comparison to the bare mechanical resonance at $\omega_m$.

While these resonance features might be useful for some applications, our interest is in the broadband sensitivity, rather than the sensitivity to monochromatic signals. With monochromatic signals, noise optimization at a specific frequency, especially efforts to tune a setup around the resonance frequencies, is important. In particular, the resonances present in the backaction noise term correspond to target frequencies for which backaction noise is completely eliminated. This strict backaction evasion is distinct from a reduction in backaction noise over a broad range of frequencies, characteristic of QND-like measurements, as we now discuss.

Of particular interest is the behavior observed in the region of frequency above the mechanical resonance but below the cavity decay rate $\kappa$. This is a consequence of our signal of interest: an impulse delivered over a very short time. Thus, we are interested in making measurements on the timescale associated with this frequency range. We note a sharp contrast in the behavior of the magnetomechanical current readout and electromechanical voltage readout cases (Figs.~\ref{fig:noise}b and \ref{fig:noise}c) when compared to that of magnetomechanical voltage readout and electromechanical current readout (Figs.~\ref{fig:noise}a and \ref{fig:noise}d) --- a consequence of the different mechanical degrees of freedom accessed in each set of cases. In the former, backaction and shot noise are constant in the regions for which $\nu < \omega_m \text{, } \delta_v$, and in the region $\omega_m \text{, } \delta_v < \nu< \kappa$, diverge as $\nu^4$. This is consistent with the behavior observed in the position-sensing case shown in Fig.~\ref{previous}. In the latter, backaction and shot noise go as $\nu^{-2}$ for $\nu < \omega_m$ and $\nu^{2}$ for $\omega_m< \nu< \kappa$. In other words, the magnetomechanical voltage and electromechanical current schemes exhibit a decrease in backaction noise in the vicinity of the mechanical resonance, analogous to the velocity-sensing case.

At high frequency where $\nu > \kappa$, we see similar behavior in the backaction and shot noise terms across all of the detector and readout combinations. In particular, backaction noise is either constant in this region, as in Figs.~\ref{fig:noise}a and \ref{fig:noise}d for magnetomechanical voltage readout and electromechanical current readout, respectively, or diverges as $\nu^2$, as in Figs.~\ref{fig:noise}b and \ref{fig:noise}c for magnetomechanical current readout and electromechanical voltage readout, respectively. On the other hand, shot noise diverges as either $\nu^4$ for magnetomechanical voltage readout and electromechanical current readout or $\nu^6$ for magnetomechanical current readout and electromechanical voltage readout.

\subsection{Analysis in the context of Windchime}
\label{windchime}

Recent advances in sensing technologies suggest that we can search for dark matter (DM) candidates through their gravitational interaction alone by building an array of many mechanical sensors~\cite{carney2020proposal}. Based on this proposal, the Windchime collaboration is developing the necessary experimental techniques and devices for the gravitational detection of DM candidates around the Planck mass range ($\sim 21.76 ~ \mu $g)~\cite{snowmass}. In particular, we are considering milligram- to gram-scale sensors with very low natural resonance frequencies (about $1$-$100$ Hz) and with significant environmental isolation using a dilution refrigerator at temperatures of $10 ~ {\rm mK}$, which makes the thermal noise floor extremely low. We are thus mostly limited by measurement-added noise. We wish to compare the performances of voltage and current readout of the magnetomechanical detector scheme to the SQL-level benchmark associated with force measurement, specifically in the context of a signal of interest for the Windchime collaboration.

For an individual sensor of mass $m$ in the array, a DM candidate of mass $m_{\rm dm}$, passing at a distance $b$ and with velocity $v$, interacts with the sensor through the Newtonian gravitational force. We are interested in the component of the gravitational force which is perpendicular to the DM candidate's trajectory. This is our intended signal~\cite{sohitri}, 
\begin{equation}
F_{\rm{sig}}(t) = \frac{G_N m m_{\rm dm} b}{(b^2+v^2 t^2)^{3/2}} \enspace .
\end{equation}
In the frequency domain, it takes the form
\begin{equation}
\label{exactsig}
F_{\rm{sig}}(\nu) = \sqrt{\frac{2}{\pi}}\frac{G_N m m_{\rm dm} |\nu|}{v^2} K_1\left(\frac{b}{v}|\nu|\right) \enspace ,
\end{equation}
where $G_N$ is the gravitational constant and $K_1$ is a modified Bessel function. We note that this signal is well approximated by
\begin{equation}
\label{approxsig}
F_{\rm{sig}}^{\rm{approx}}(\nu) = \sqrt{\frac{2}{\pi}} \frac{G_N m m_{\rm dm}}{b v} e^{-b |\nu|/2 v} \enspace ,
\end{equation}
which we use to estimate a signal-to-noise ratio (SNR).

The signal is delivered over a very short period of time, set by the timescale $\tau \sim b/v$. In the frequency domain, this translates to a constant broadband signal that rapidly diminishes at a frequency set by the timescale $\tau$, thereby determining the signal bandwidth.
For example, if we consider a DM candidate with a velocity around $200 ~ {\rm km/s}$~\cite{DMvelocity} passing at a distance $b \sim 1 ~ {\rm mm}$, the timescale of the signal is approximately $\tau \sim 10^{-8}$~s. In the frequency domain, this signal is constant until very high frequency, approximately $1$ GHz, after which the signal falls off to zero. We wish to identify this signal amidst collected time series data. As  explored in Ref.~\cite{sohitri}, an efficient search strategy for such a broadband signal is to use an optimal filter to scan through the time series data, yielding the SNR
\begin{equation}
\label{SNR}
\text{SNR}^2_{\text{opt}} =\int_{0}^{\infty} \frac{|F_{\rm sig}(\nu)|^2}{S_{\rm FF}(\nu)} d\nu \enspace ,
\end{equation}
where the effective bandwidth of integration is set by the signal's timescale.

\begin{figure}
\begin{center}
\includegraphics[width=0.48 \textwidth]{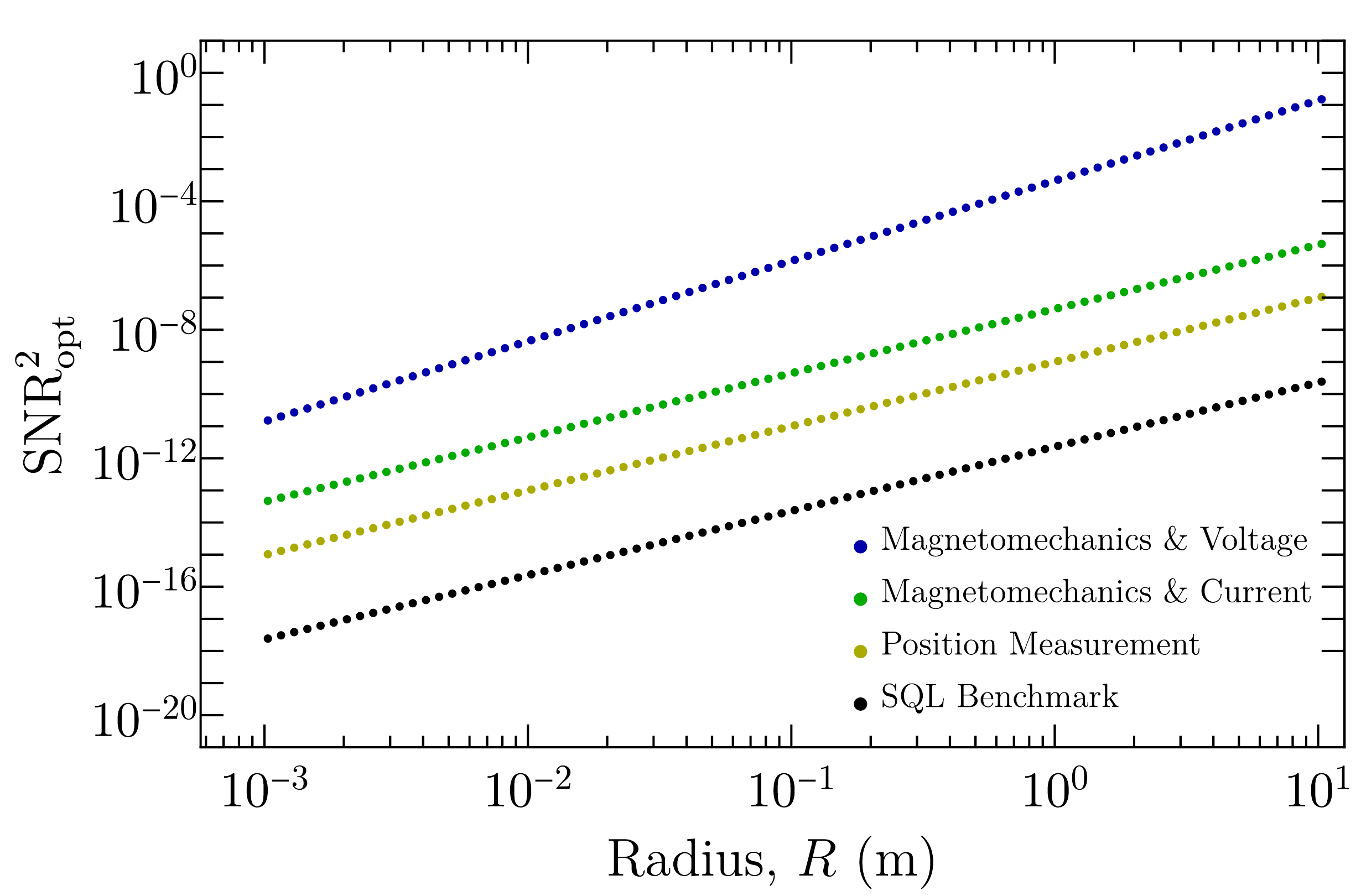}
\caption{The SNR$^2_{\rm opt}$ plotted as a function of the radius of the sensor, starting from an original radius of $R \approx 1$ mm and the parameters listed in Figs.~\ref{fig:comparison} and \ref{fig:noise} for the magnetomechanical detector scheme. The mass scales according to $m = \rho \pi R^2 h$, where we fix the mass density to $\rho = 7500$ kg/m$^3$ and allow the height $h$ to scale linearly with $R$ to maintain a fixed height to radius ratio of approximately $40$. The transducer constant $T_v$ and inductance of the voice coil $L$ scale according to Eq.~(\ref{eq:voltage}) and $L = \frac{\mu_0 N^2 \pi R^2}{h}$, respectively, where the total turn number scales according to $N = n h$ and we fix the turn density to $n \approx 7725$ turns/m and the magnetic field to $B = 1$~T. We fix the characteristic impedance of each LC circuit to $Z_0^v = 2 \pi \cdot 10^2$ $\Omega$ for voltage readout and $Z_0^x \approx 2 \pi$~$\Omega$ for current readout, maintaining the ratio $L/L_M = 10^4$ and forcing the capacitance $C_L$ to scale with R. As the optimized coupling strengths given by Eqs.~(\ref{eq:GQvSQL}) and (\ref{eq:GBvSQL}) depend on these scaling parameters, $G_Q^{(E,v)}$ and $G^B$ also scale with $R$ where we opt to scale the target frequency as $R^{-1}$ from the original 1 MHz. The same scaling applies to the SNR$^2_{\rm opt}$ for position measurement (yellow), using the noise PSD and initial parameters indicated in Fig.~\ref{previous} where we neglect damping and scale the target frequency down from $0.1$ MHz.}
\label{fig:SNR-square}
\end{center}
\end{figure}

We estimate the SNR using the approximated signal in Eq.~(\ref{approxsig}) and the noise PSDs appropriate for voltage or current readout of the magnetomechanical detector scheme, given by Eqs.~(\ref{magnetic-voltage}) and (\ref{magnetic-current}), respectively. In Fig.~\ref{fig:SNR-square}, we show the SNRs as a function of the radius $R$ of the cylindrical test mass (i.e., the sensor of mass $m$). Scaling the radius impacts a variety of parameters and circuit quantities, including the mass of the magnetic sensor, the inductance of the voice coil, and the transducer constant, as increasing the size of the magnetic mass requires the voice coil to scale up as well. Relevant details can be found in the caption of Fig.~\ref{fig:SNR-square}. We note that we take the distance $b$ to scale linearly with $R$, which yields a timescale $\tau$ that inherently depends on $R$. Therefore, in calculating the SNR in Eq.~(\ref{SNR}), the effective bandwidth of integration is implicitly set by the size of the mass.

We compare these SNRs with that of the SQL-level noise floor associated with a force measurement~\cite{caves1980measurement}, in which we infer the force acting on the sensor by monitoring the position of a free particle over time, where the measurements are separated by time $\tau$. This is given by the relation
\begin{equation}
\Delta F_{\rm SQL} \sim \sqrt{\frac{\hbar m_s}{ \tau^3}} \enspace .
\end{equation}
In Fig.~\ref{fig:SNR-square}, we plot the ratio $[ F_{\rm sig}(\tau)/\Delta F_{\rm SQL} ]^2$ as a function of $R$. For additional comparison, we include the SNR$^2_{\rm opt}$ corresponding to the position-sensing noise PSD of an optomechanical system shown in Fig.~\ref{previous} and described in Appendix~\ref{Previous-Work}.

We find the magnetomechanical detection scheme with either readout option offers an improved sensitivity over both the SQL benchmark and standard optomechanical position sensing. In particular, voltage readout demonstrates orders of magnitude improvement in the SNR. For example, at a radius of approximately $10$~cm, we observe about a 39~dB improvement over the SQL benchmark and a 26~dB improvement relative to standard position sensing. In contrast, current readout of the magnetomechanical detector scheme only offers about a 21~dB improvement over the SQL benchmark and a 8~dB improvement over standard position sensing. In addition, we see an overall improved sensitivity as the size of the test mass is scaled up, with the SQL benchmark, position-sensing case, and current readout of the magnetomechanical detector scheme scaling as $R^2$ and voltage readout of the magnetomechanical detector scheme scaling as $R^{2.5}$. We attribute the difference in scaling between the two readout options to be a consequence of the distinct circuits associated with each scheme, resulting in unique noise PSDs each with a different dependence on $R$. While these are encouraging results, we caution that these improvements indicate a $10$~cm-radius sensor requires a test mass of $10^3$~kg (1 metric ton) in order to measure a voltage signal on the order of attovolts. Granted, these estimates may be improved by considering instead the density and magnetic field of a superconducting material, rather than a strong permanent magnet (neodymium), as we have~here.

\section{Outlook}
\label{Conclusions}

Here we develop specific approaches for velocity and position sensing using voltage or current measurements of magnetomechanical and electromechanical transducers. We find that our specific electrical circuit-based approach to velocity sensing, namely voltage measurement of a magnetomechanical transducer, allows for a reduction in measurement-added noise while monitoring the mechanical motion in the microwave domain. 
While it is well known that Faraday's law connects voltage and velocity, this has not been used as a method for velocity measurement to date. Here we have shown that this may be a very fruitful domain for future exploration that is immediately compatible with existing mechanical systems, such as levitated superconducting spheres (e.g., the ones described in Refs.~\cite{gerard,gerard2}), by effectively changing their motional readout from current to voltage.

Applying the approaches we describe here to the challenge of direct dark matter detection showcases how this type of readout enables scaling to very large masses while keeping a very low floor of quantum noise.
In this work, using very large objects is advantageous for measuring gravitational signals, as we focus primarily on observing small accelerations.
In contrast, the observation of small forces requires a very different operating regime, offering an intriguing prospect for future work: the design of an effective small force sensor.

This type of velocity measurement is a QND-like measurement --- a consequence of the QND structure of the velocity variable in the context of a mechanical oscillator well above its resonance frequency, i.e., in the free particle limit. Consequently, we anticipate that this could be a critical choice to make for future systems that incorporate sensors that need to operate in an impulse-sensing domain.
Furthermore, our simple implementation of voltage measurement for a magnetomechanical transducer, which is a variation on the well-known rf-SET, can likely be improved with modern circuit QED techniques.

\begin{acknowledgements}

We thank A.~A.~Clerk, J.~Teufel, A.~Chou, and R.~F.~Lang for helpful discussions. Work at LBL is supported by the U.S. DOE, Office of High Energy Physics, under Contract No. DEAC02-05CH11231 and the Quantum Information Science Enabled Discovery (QuantISED) for High Energy Physics Grant KA2401032. This work was supported by the U.S. DOE Office of Science, Office of High Energy Physics, QuantISED program (under FWP ERKAP63) as well as the Swedish Research Council under Grant 2020-00381 (G.H.). This work is also supported by the Department of Energy through the Fermilab Theory QuantiSED program in the area of “Intersections of QIS and Theoretical Particle Physics". Fermilab is operated by the Fermi Research Alliance, LLC under Contract DE-AC02-07CH11359 with the U.S. Department of Energy.

\end{acknowledgements}

\appendix

\section{Faraday's law}
\label{faraday}

In the magnetomechanical detection scheme, our interest lies in the voltage generated when the magnetic mass moves through the stationary voice coil as a result of some impulse. This induced voltage $\varepsilon$ is given by Faraday's law~\cite{jackson,zangwill}. Faraday's law relates $\varepsilon$ to the time derivative of the magnetic flux $\Phi_B = \int_S \bold{B} \cdot \bold{dA}$ through an open surface $S$: 
\begin{equation}
\label{eq:FL}
\varepsilon = -\frac{d \Phi_B}{dt} = -\frac{d}{dt} \left( \int_S \bold{B} \cdot \bold{dA} \right) \enspace .
\end{equation}
However, in this case it is advantageous to re-express Eq.~(\ref{eq:FL}) in terms of two contributions, both integrated around the closed path $C$ that bounds the surface $S$:
\begin{equation}
\label{eq:FLexp}
\varepsilon = \oint_C \bold{E} \cdot \bold{d \ell} + \oint_C (\bold{v \times B}) \cdot \bold{d \ell} \enspace .
\end{equation}
The first term accounts for the electric field $\bold{E}$ generated by a time-varying magnetic field, while the second term accounts for time-varying changes in $S$ and $C$, i.e., motion of the curve with some velocity $\bold{v}$. By considering the rest frame of the magnetic mass and taking the coil to be moving with a velocity $\bold{v} = v \bold{\hat{z}}$, Eq.~(\ref{eq:FLexp}) can be evaluated to yield the induced voltage given by Eq.~(\ref{eq:voltage}) in the main text.

By expressing Faraday's law as in Eq.~(\ref{eq:FLexp}) we more readily understand the equivalence between the rest frames of the magnetic mass and voice coil. In the rest frame of the coil, the velocity is zero, however, a time-varying magnetic field will be present due to the magnet's motion, yielding a non-zero electric field. We use a transformation between rest frames \cite{jackson,zangwill} to express this electric field in terms of the magnetic field in the magnet's rest frame: $\bold{E} = \bold{v \times B}$. In this way, we see how the induced voltage in our detection scheme is equivalent to that given by Eq.~(\ref{eq:voltage}).

\section{Circuit analysis of the detector schemes}
\label{circuits101}

In this appendix, we provide a pedagogical presentation of circuit quantization techniques (specifically the node flux method, following Ref.~\cite{devoret}) as applied to the detector schemes outlined in Section~\ref{S1} of the main text. This connects the circuit degrees of freedom to those of the mechanical systems we consider to arrive at the Hamiltonians in Eqs.~(\ref{eq:H1mV}), (\ref{eq:H2mV}), and (\ref{eq:Hem}).

In an electrical circuit, every circuit element is characterized by a branch voltage and a branch current, whose time integral defines the element's branch flux and branch charge, respectively:
\begin{equation}
\label{eq:bvar}
\begin{split}
\Phi^b (t) = \int^t_{-\infty} v^b (t') dt' \enspace , \\
Q^b (t) = \int^t_{-\infty} i^b (t') dt' \enspace .
\end{split} 
\end{equation}
We use the superscript $b$ to denote branch variables throughout the text; this is to distinguish from other variables, such as velocity $v$, present in our analysis. Circuit elements are characterized by fundamental equations that relate their branch current or charge to their branch voltage or flux. For example, in capacitors, $v^b = \dot{\Phi}^b = Q^b/C$, while for inductors $i^b = \dot{Q}^b = \Phi^b/L$. A nonlinear element such as a Josephson junction is characterized by the relation $i^b = \dot{Q}^b = I_C \sin(2 \pi \Phi^b / \Phi_0) + C_J \ddot{\Phi}^b$, with $I_C$ its critical current, $C_J$ its self-capacitance, and $\Phi_0$ the magentic flux quantum.

Kirchhoff's laws determine how the branch variables of each element of a circuit relate. Kirchhoff's current law enforces charge conservation at each node by equating the currents flowing into and out of each node. Kirchhoff's voltage law (an instance of Faraday's law) demands that the voltage around a closed loop must sum to zero. However, the branch variables do not constitute the degrees of freedom of the circuit, as they are not independent variables. In order to appropriately define independent degrees of freedom for a circuit, a so-called `spanning tree'~\cite{devoret} must be chosen. The spanning tree determines how each of the branch variables may be expressed in terms of the defined independent degrees of freedom.

In the node flux formulation, a spanning tree is constructed as follows. Starting from a designated ground or reference node, a path is chosen along each branch element such that each non-reference node of the circuit is reached by a single path. Each non-reference node is then associated with a node flux, defined as the sum (or difference) of the branch fluxes along the path to each node.
The various possible spanning trees for a given circuit amount to different gauge choices and are therefore distinct but equivalent descriptions. Once expressed in terms of node fluxes, the set of equations found from Kirchhoff's current law become the equations of motion for each degree of freedom, which are used to then infer the Lagrangian of the system. With this method, capacitive terms yield time derivatives of the node flux, thereby playing the role of kinetic energy terms. In contrast, inductive terms are written in terms of the node flux, hence acting like potential energy terms. In this way, Kirchhoff's current laws become equations of motion, while the Kirchhoff's voltage laws determine how the branch fluxes are defined in terms of the node fluxes.

We use this general procedure to find the Lagrangians and derive the Hamiltonians for each of the detector configurations presented in the main text. In the magnetomechanical case (Section~\ref{s1mm}), we consider the lumped-element circuit shown in Fig.~\ref{fig:visual}b. We highlight two distinct yet equivalent gauge descriptions, which when combined with the mechanics of the system, reveal different couplings between the circuit and mechanical degrees of freedom. In the electromechanical case (Section~\ref{s1em}), we examine the lumped-element circuit shown in Fig.~\ref{fig:visual}d, where we find a nonlinear coupling between the circuit and mechanical degrees of freedom. We then perform an expansion about the minimum energy configuration to linearize this coupling.

\subsection{Magnetomechanical circuit analysis}

For the magnetomechanical case, we consider the detector circuit shown in Fig.~\ref{fig:visual}b of the main text. Upon choosing the bottom node as the ground node, we write the equations found using Kirchhoff's laws~as
\begin{equation}
\label{eq:KL}
\begin{split}
C_L \ddot{\Phi}^b_{C_L} &= \frac{\Phi^b_L}{L} \\
\dot{\Phi}^b_L + \dot{\Phi}^{\text{ext}} &= -\dot{\Phi}^b_{C_L} \enspace ,
\end{split}
\end{equation}
where we have chosen an orientation of the coil relative to the magnet such that the self-induced voltage and externally generated voltage are additive. Note that the inductor's voltage must reflect both its own contribution (the self-induced voltage) as well as that from the external flux of the magnet, namely, its time-derivative, given that $\Phi^{\text{ext}}$ is generally a time-dependent quantity.

\begin{figure}
\includegraphics[width=0.48 \textwidth]{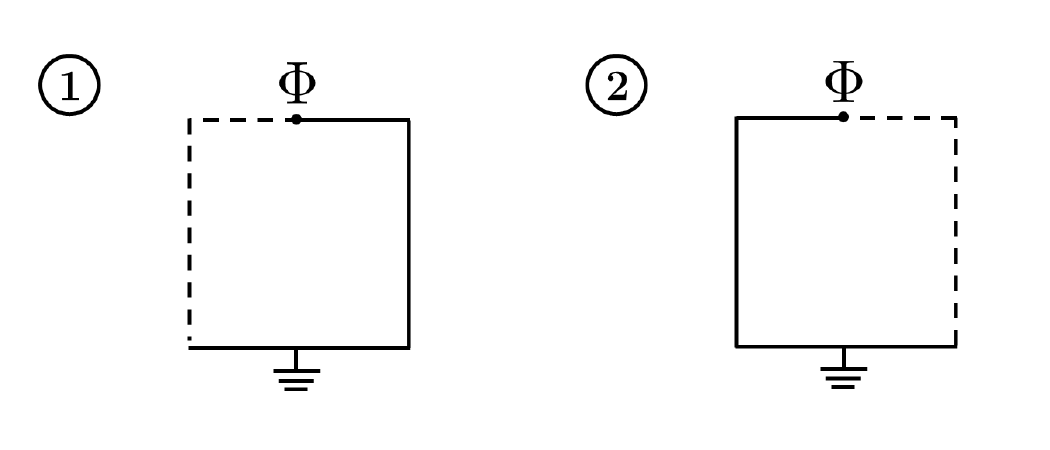}
\caption{Schematic highlighting the two spanning trees associated with each gauge in the magnetomechanical detector configuration.}
\label{fig:span}
\end{figure}

In this case, there is only one non-reference node and only two potential paths to reach it: through the capacitor or through the inductor.
For each of these spanning trees, shown in Fig.~\ref{fig:span}, we define an appropriate node flux. The presence of an external flux results in two spanning trees with distinct definitions of each branch flux in term of the node flux. For the first path (through the inductor), we define $\Phi = \Phi^b_L$. Alternatively, for the second path (through the capacitor) we find $\Phi = -\Phi^b_{C_L}$. By implementing these distinct node flux definitions in Eq.(\ref{eq:KL}), we see that the voltage law defines each alternate branch flux in terms of the node flux, and the current law provides the equation of motion governing $\Phi$.

In the first case, where $\Phi = \Phi^b_L$, Eq.(\ref{eq:KL}) implies that $\dot{\Phi}^b_{C_L} = -(\dot{\Phi} + \dot{\Phi}^{\text{ext}})$ and yields the equation of motion
\begin{equation}
\label{eq:C1}
C_L (\ddot{\Phi} + \ddot{\Phi}^{\text{ext}}) = -\frac{\Phi}{L}  \enspace .
\end{equation}
In the second case, where $\Phi = -\Phi^b_{C_L}$, Eq.(\ref{eq:KL}) implies that $\dot{\Phi}^b_L = \dot{\Phi} - \dot{\Phi}^{\text{ext}}$ and yields the equation of motion
\begin{equation}
\label{eq:C2}
C_L \ddot{\Phi} = -\frac{(\Phi-\Phi^{\text{ext}})}{L} \enspace .
\end{equation}
These equations of motion can be used to obtain the Lagrangians, which for each respective spanning tree are written as
\begin{equation}
\label{eq:L1}
\mathcal{L}_1 = \frac{1}{2} C_L (\dot{\Phi} + \dot{\Phi}^{\text{ext}})^2 - \frac{\Phi^2}{2L}  
\end{equation}
and
\begin{equation}
\label{eq:L2}
\mathcal{L}_2 = \frac{1}{2} C_L \dot{\Phi}^2 - \frac{(\Phi - \Phi^{\text{ext}})^2}{2L} \enspace .
\end{equation}
With the Lagrangians of each spanning tree specified, we use the usual Legendre transformation to obtain the Hamiltonians~\cite{gold} in each case: 
\begin{equation}
\label{eq:H1}
H_1 = \frac{Q^2}{2 C_L} + \frac{\Phi^2}{2L} - Q \dot{\Phi}^{\text{ext}}  
\end{equation}
and
\begin{equation}
\label{eq:H2}
H_2 = \frac{Q^2}{2 C_L} + \frac{(\Phi - \Phi^{\text{ext}})^2}{2L} \enspace ,
\end{equation}
where $Q = \frac{d\mathcal{L}_i}{d\dot{\Phi}}$ represents the canonical charge degree of freedom, conjugate to the node flux $\Phi$. With only the lone capacitance $C_L$ connected to the node, $Q$ represents the charge on this capacitor's plates. These Hamiltonians are related by the gauge transformation given by Eq.~(\ref{eq:GT}) in Section~\ref{s1mm} of the main text.

To incorporate the mechanical degrees of freedom due to the magnetic mass's motion, we recall that $\Phi^{\text{ext}}$ represents the flux penetrating the voice coil due to the presence of the magnetic mass. Therefore, $\dot{\Phi}^{\text{ext}}$ corresponds to the induced voltage $\varepsilon$ in Eq.~(\ref{eq:voltage}) of the main text. This enables us to rewrite Eq.~(\ref{eq:voltage}) in terms of $\Phi^{\text{ext}}$, the mechanical degrees of freedom, and the transducer constant~$T_v$:
\begin{equation}
\label{eq:phimech}
\begin{split}
\dot{\Phi}^{\text{ext}} &= T_v \dot{x} \\
\Phi^{\text{ext}} &= T_v x \enspace ,
\end{split}
\end{equation}
where $x$ represents the mass's position and $\dot{x}$ its velocity $v$. We then incorporate the motion of the mass and its attached spring in the Lagrangians of Eqs.~(\ref{eq:L1}) and (\ref{eq:L2}) by making the substitutions indicated in Eq.~(\ref{eq:phimech}) and including terms that describe the energy associated with the mechanical motion. In full, we come to the Lagrangians
\begin{equation}
\label{eq:L1m}
\mathcal{L}_1^{(E,v)} = \frac{1}{2} m \dot{x}^2 - \frac{1}{2} k x^2 + \frac{1}{2} C_L (\dot{\Phi} + T_v \dot{x})^2 - \frac{\Phi^2}{2L}  
\end{equation}
and
\begin{equation}
\label{eq:L2m}
\mathcal{L}_2^{(E,v)} = \frac{1}{2} m \dot{x}^2 - \frac{1}{2} k x^2 + \frac{1}{2} C_L \dot{\Phi}^2 - \frac{(\Phi - T_v x)^2}{2L} \enspace .
\end{equation}
Moving to the Hamiltonian description via a Legendre transform yields the Hamiltonians given in Eqs.~(\ref{eq:H1mV}) and~(\ref{eq:H2mV}).

As an aside, we can confirm the Lagrangians in Eqs.~(\ref{eq:L1m}) and (\ref{eq:L2m}) appropriately characterize the mechanics by considering the forces acting on the magnetic mass. Theses forces include the restorative force of the spring as well as a magnetic force due to the interaction between the current-carrying voice coil and the magnetic mass. Due to the interaction between the current in the voice coil and the magnetic field in the air gap of the magnetic mass, the voice coil experiences a magnetic force of the form $F = \int i^b_L \bold{d \ell} \cross \bold{B} = - T_v i^b_L$. As a result of Newton's third law, the force felt by the magnetic mass is equal and opposite to the force felt by the voice coil. The equation of motion for the position of the magnetic mass can then be written as
\begin{equation}
m \ddot{x} = -kx + \frac{T_v}{L} \Phi^b_L \enspace ,
\end{equation} 
where we have made the substitution for the branch flux of the inductor via $i^b_L = \Phi^b_L / L$. One can confirm that by making the appropriate substitutions for branch flux in each gauge yields an equation of motion (combined with either Eq.~(\ref{eq:C1}) or Eq.~(\ref{eq:C2})) generated by the Lagrangians in either Eq.~(\ref{eq:L1m}) or Eq.~(\ref{eq:L2m}).

\subsection{Electromechanical circuit analysis}

For the electromechanical configuration, we begin with the lumped-element circuit shown in Fig.~\ref{fig:visual}d of the main text and designate the bottom node as the ground or reference node and assemble the appropriate expressions from Kirchhoff's laws. This yields the equations 
\begin{equation}
\label{eq:EMKL}
\begin{split}
\frac{d}{dt} \left[ C(x) \dot{\Phi}^b_C \right] &= \frac{\Phi^b_L}{L} + C_P \ddot{\Phi}^b_{C_P} \enspace ,\\ 
V_{\text{DC}} + &\dot{\Phi}^b_C + \dot{\Phi}^b_L = 0 \enspace ,\\
\dot{\Phi}^b_L &= \dot{\Phi}^b_{C_P} \enspace ,
\end{split}
\end{equation}
where the first equation arises from applying Kirchhoff's current law to the top node and the remaining equations are the result of applying Kirchhoff's voltage law around each of the circuit's two loops. Note that we have written the current through the mechanically-varying capacitor generally in terms of the total time derivative of the charge on the capacitor's plates $Q^b_C = C(x) \dot{\Phi}^b_C$. This is due to the fact that the capacitance is a function of the mechanical position.

We next construct a spanning tree to define the node flux $\Phi$. While there are three potential spanning trees in this circuit, without any externally threaded flux all three choices yield identical definitions of the branch fluxes in terms of the node flux. Namely, $\dot{\Phi}^b_C = \dot{\Phi} - V_{DC}$ and $\dot{\Phi}^b_L = \dot{\Phi}^b_{C_P} = -\dot{\Phi}$. Expressing Kirchhoff's current law in Eq.~(\ref{eq:EMKL}) in terms of the node flux, we find the equation of motion 
\begin{equation}
\label{eq:EMeom}
\frac{d}{dt} \left[ C(x) \left(\dot{\Phi} - V_{\text{DC}} \right) \right] = -\frac{\Phi}{L} - C_P \ddot{\Phi} \enspace .
\end{equation}
Working backwards, we infer the Lagrangian that describes the circuit dynamics:
\begin{equation}
\label{eq:EML}
\mathcal{L} = \frac{1}{2} C(x) \left(\dot{\Phi} - V_{\text{DC}} \right) ^2 + \frac{1}{2} C_P \dot{\Phi}^2  - \frac{\Phi^2}{2 L} \enspace.
\end{equation}
This Lagrangian does not fully describe the system as it does not completely account for the mechanical motion. However, we need only add the usual mechanical contributions due to the kinetic energy of the plate and the potential energy of the attached spring, yielding the total Lagrangian
\begin{equation}
\label{eq:EML}
\begin{split}
\mathcal{L}^{(E,x)} = &\frac{1}{2} m \dot{x}^2 - \frac{1}{2} k x^2 + \frac{1}{2} C(x) \left(\dot{\Phi} - V_{\text{DC}} \right) ^2 \\
&+ \frac{1}{2} C_P \dot{\Phi}^2  - \frac{\Phi^2}{2 L}  \enspace .
\end{split}
\end{equation}

We can confirm that this Lagrangian appropriately accounts for the mechanical degrees of freedom by considering the forces acting on the movable plate: the restorative force of the attached spring and the electrostatic attraction between the oppositely charged plates of the capacitor. When the plates are uncharged, we take the spring to be in its equilibrium position so that the plate's position is $x=0$ and the plate separation is $d_0$. In this coordinate system, we express the mechanically-varying capacitance as $C(x) = \frac{\epsilon_0 A}{d_0-x}$ where $A$ is the area of the two plates and $\epsilon_0$ the permittivity of free space. Assuming that the area of plates is much larger than their original separation such that they may be treated approximately as two infinite sheets with charge $\pm Q^b_C$, the electric field between them is uniform, given by $E = \frac{Q^b_C}{\epsilon_0 A}$, and the force of attraction felt by the movable plate is given by $F = \frac{1}{2}E Q^b_C = \frac{(Q^b_C)^2}{2 \epsilon_0 A}$. Using Newton's second law, the equation of motion for the position of the movable plate is then
\begin{equation}
\label{eq:EMmecheom}
\begin{split}
m \ddot{x} &= -k x + \frac{(Q^b_C)^2}{2 \epsilon_0 A} \\
&= -k x + \frac{C(x)^2 (\dot{\Phi} - V_{\text{DC}})^2}{2 \epsilon_0 A} \\
&= -k x + \frac{1}{2} \frac{\partial C(x)}{\partial x} (\dot{\Phi} - V_{\text{DC}})^2 \enspace ,
\end{split}
\end{equation}
where in the second line we have made the substitution $Q^b_C = C(x)\dot{\Phi}^b_C = C(x)(\dot{\Phi} - V_{\text{DC}})$ such that the mechanical equation of motion is expressed in terms of the circuit's degree of freedom, the node flux $\Phi$. This is further simplified in the third line by noting $\frac{\partial C(x)}{\partial x} = \frac{\epsilon_0 A}{(d_0-x)^2} = \frac{C(x)^2}{\epsilon_0 A}$. Working backwards, we can confirm that this equation of motion for $x$ is generated by the Lagrangian in Eq.~(\ref{eq:EML}).

We then use this Lagrangian and the usual Legendre transformation~\cite{gold} to find the Hamiltonian: 
\begin{equation}
\label{eq:EMfullH}
H^{(E,x)} = \frac{p^2}{2m} + \frac{\Phi^2}{2L} + V(Q,x) \enspace ,
\end{equation}
where we define the quantity $V(Q,x)$ as
\begin{equation}
\label{eq:Vem}
V(Q,x) = \frac{1}{2} k x^2 + \frac{Q^2 + 2 C(x) V_{\text{DC}} Q - C_P C(x) V_{\text{DC}}^2}{2(C(x) + C_P)}
\end{equation}
and identify $Q$ and $p$ as the canonical node charge and momenta conjugate to the node flux $\Phi$ and position $x$, respectively. In this case, $Q$ corresponds to the sum of the charge on the plates of the two capacitors connected to the node, while $p$ is the mechanical momentum of the movable plate. The coupling between the circuit and the mechanical degrees of freedom is contained in $V(Q,x)$ and unsurprisingly, this coupling is nontrivial since the capacitance $C(x)$ is inversely proportional to $x$. However, by considering small displacements of the movable plate of the capacitor, we can linearize the Hamiltonian by expanding about the equilibrium of the circuit and mechanical systems.

We understand this equilibrium as follows. When the plates are uncharged, the separation between them is given by $d_0$ and the location of the movable plate is at $x=0$. Once charged, the electrostatic attraction between the oppositely charged plates brings them closer together, resulting in a new position for the movable plate at $x = x_0$. This position corresponds to the position where the restorative force of the spring and the force of electrostatic attraction are balanced. Examining the Hamiltonian, the contribution $V(Q,x)$ describes a two-dimensional potential energy landscape dependent on charge and position in which this point must be minimum. This equilibrium is where the plate exists upon perturbation due to some detection event. If we assume these perturbations are small, we can approximate this two-dimensional landscape by considering just the region  in the vicinity of the equilibrium $\lbrace Q_0,x_0 \rbrace$, thereby linearizing the charge-position interaction. Therefore, we seek the solution $ \lbrace Q_0,x_0 \rbrace $ to the equation $\nabla V(Q,x) = 0$ such that $ \lbrace Q_0,x_0 \rbrace $ is a minimum.

With this equilibrium point in hand, we expand $V(Q,x)$ about this point to find
\begin{equation}
\label{eq:Vexp}
\begin{split}
V(Q,x) = &V(Q_0,x_0) + \nabla V(Q_0,x_0) \cdot \underline{Q} \\
&+ \frac{1}{2} \underline{Q}^T  \underline{\underline{D}} \ \underline{Q}  + ... \enspace ,
\end{split}
\end{equation}
where we have defined the vector $ \underline{Q} = \lbrace Q-Q_0, x-x_0 \rbrace$ and the matrix $\underline{\underline{D}}$ as the Hessian matrix evaluated at the minimum $ \lbrace Q_0,x_0 \rbrace $. We express $\underline{\underline{D}}$ in the compact form
\begin{equation}
\label{eq:Dmat}
\underline{\underline{D}} = \begin{bmatrix}
\frac{1}{C_{\text{eff}}} & -\frac{T_x}{C_P} \\[0.6em]
-\frac{T_x}{C_P} & k_{\text{eff}}
\end{bmatrix}
\end{equation}
where have defined the effective capacitance $C_{\text{eff}} = C(x_0) + C_P$, the effective spring constant $k_{\text{eff}} = k - \frac{C_{\text{eff}} T_x^2}{C(x_0) C_P}$, and the transducer constant $T_x$ as
\begin{equation}
T_x = C_P \frac{\left( Q_0-C_P V_{\text{DC}} \right)}{\epsilon_0 A} \left( \frac{C(x_0)}{C_{\text{eff}}} \right)^2 \enspace .
\end{equation}
We understand this transducer constant to be the constant of proportionality that takes changes in the position of the movable plate to changes in the charge on the capacitor's plates. That is, if we examine the expression for the charge on the mechanically-varying capacitor's plates and expand about the minimum $ \lbrace Q_0,x_0 \rbrace $,
\begin{equation}
\begin{split}
Q^b_C &= \frac{C(x)\left( Q-C_P V_{\text{DC}} \right)}{C(x) + C_P} \\
&\approx \frac{C(x_0)}{C_{\text{eff}}} (Q - C_P V_{\text{DC}}) + T_x (x-x_0) \ + ... \enspace ,
\end{split}
\end{equation}
we find the approximately linear relationship between the movable plate's position and the charge on the mechanically-varying capacitor's plates via the transducer constant~$T_x$.

Inserting the expansion of $V(Q,x)$ from Eq.~(\ref{eq:Vexp}) into the Hamiltonian of Eq.~(\ref{eq:EMfullH}) (noting that the linear order term is zero at the minimum), the expanded linearized Hamiltonian then takes the form given by Eq.~(\ref{eq:Hem}) in the main text.

\section{Adding the drive and moving to the linearized regime}
\label{optomechanics}

In this appendix, we provide a brief overview of the standard methods employed in optomechanical analyses~\cite{bowen,clerk,aspel} (as applied to our systems of interest) that lead to Eqs.~(\ref{eq:HVfinalmm})-(\ref{eq:HIfinalem}) in Section~\ref{S2} of the main text. We begin by coupling the schemes for electric and magnetic field sensing to a bath that serves as a source of drive and mode of dissipation. Thus, we include in the Hamiltonians the terms
\begin{equation}
\label{eq:Hbath}
\hat{H}_B = \int^\infty_{-\infty} d\omega \hbar \omega \hat{b}^\dagger(\omega) \hat{b}(\omega) \enspace ,
\end{equation}
and
\begin{equation}
\label{eq:HBI}
\hat{H}_{\text{int}} = i \hbar \sqrt{\frac{\kappa}{2\pi}} \int^\infty_{-\infty} d\omega \left[\hat{b}^\dagger(\omega)\hat{a} - \hat{b}(\omega) \hat{a}^\dagger\right] \enspace ,
\end{equation}
where $\hat{b}^\dagger(\omega),\hat{b}(\omega)$ are the creation and annihilation operators for the bath modes, which satisfy the commutation relation $[\hat{b}(\omega),\hat{b}^\dagger(\omega')]=\delta(\omega-\omega')$, and $\kappa$ corresponds to the cavity decay rate. Eq.~(\ref{eq:Hbath}) represents the Hamiltonian of the bath while Eq.~(\ref{eq:HBI}) is the coupling between the parametric cavity and the bath.

Upon coherently driving the cavity, the cavity modes are displaced from their average value such that $\hat{a} \rightarrow (\alpha + \delta\hat{a})e^{i \omega_L t}$, where $\alpha$ represents the drive strength, $\delta \hat{a}$ the operators corresponding to the dynamical quantum fluctuations about the average, and $\omega_L$ the frequency of the drive. It is then convenient to move to a frame rotating with the drive via the unitary transformation
\begin{equation}
\hat{H} \rightarrow \hat{H'} = \hat{U} \hat{H} \hat{U}^\dagger + i \hbar \frac{d\hat{U}}{dt} \hat{U}^\dagger 
\end{equation}
with 
\begin{equation}
\hat{U} = e^{i \omega_L \hat{a}^\dagger \hat{a}  t} \enspace .
\end{equation}
This unitary transformation serves to eliminate the time-dependence from the Hamiltonian, namely, in the cavity-bath interaction term $\hat{H}_{\rm int}$, as well as introduce a term $-\hbar \omega_L \hat{a}^\dagger \hat{a}$.

Finally, we move to the linearized regime of optomechanics and assume a strong drive such that we can linearize the interaction between the parametric cavity and the circuits for both voltage and current measurement. For a strong drive, the drive strength $\alpha$ increases in magnitude while also increasing the fluctuations associated with the operators $\delta\hat{a}, \delta\hat{a}^\dagger$. Thus, in the cavity-circuit coupling term (across all detector schemes, readout options, and gauges) we neglect the term going as $\delta\hat{a}^\dagger \delta\hat{a}$ as being a factor smaller in $\alpha$ than the terms $\alpha^* \delta\hat{a} + \alpha \delta\hat{a}^\dagger$. We also neglect contributions which do not dynamically affect the evolution of the system, namely, constant terms and terms linear in system operators.

By choosing the drive strength $\alpha$ to be real, and letting $\delta\hat{a}, \delta\hat{a}^\dagger \rightarrow \hat{a}, \hat{a}^\dagger$ (for convenience) we arrive at the Hamiltonians given by Eqs.~(\ref{eq:HVfinalmm})-(\ref{eq:HIfinalem}) in Section~\ref{S2} of the main~text.

\section{Defining the input and output bath modes}
\label{IOtheory}

Here we provide a review of input-output theory~\cite{colgard} to establish the quantum Langevin equation and define the input and output modes for the system. We begin with the Heisenberg equation of motion for the bath modes $\hat{b}(\omega)$. This equation is identical across all detector schemes, readout options, and gauges, 
\begin{equation}
\frac{d\hat{b}(\omega,t)}{dt} = \frac{-i}{\hbar} \ \left[\hat{b}(\omega), \hat{H}'\right] = -i \omega \hat{b}(\omega) + \sqrt{\frac{\kappa}{2\pi}} \hat{a} \enspace ,
\end{equation}
and may be solved in reference to either an initial time or final time. This solution is written as
\begin{equation}
\hat{b}(\omega,t) = e^{-i \omega (t-t_0)} \hat{b}(\omega,t_0) + \sqrt{\frac{\kappa}{2\pi}} \int_{t_0}^t dt' e^{-i \omega (t-t')} \hat{a}(t') \enspace ,
\end{equation}	
where for times $t>t_0$ the solution references an initial state at time $t_0$ and for times $t<t_0$ the solution references a final state at time $t_0$.

We then substitute this solution into the Heisenberg equations of motion for $\hat{a},\hat{a}^\dagger$ to find the quantum Langevin equation. In doing so, we define the input and output modes as
\begin{equation}
\frac{1}{\sqrt{2\pi}}\int_{-\infty}^\infty d\omega \ e^{-i\omega(t-t_0)} \hat{b}(\omega,t_0) =
\begin{cases}
    \hat{b}_{\text{in}}(t)  & \text{if } t>t_0\\
    \hat{b}_{\text{out}}(t) & \text{if } t<t_0
\end{cases}
\end{equation}
and note the identities
\begin{equation}
\int_{-\infty}^\infty d\omega e^{\pm i \omega (t-t')} = 2 \pi \delta(t-t')
\end{equation}
and
\begin{equation}
\int_a^b dx f(x) \delta(x-a) = 
\begin{cases}
    \frac{f(a)}{2}  & \text{if } b>a\\
    \frac{-f(a)}{2} & \text{if } b<a
\end{cases} \enspace .
\end{equation}
We note that the equations of motion for $\hat{a},\hat{a}^\dagger$ are unique to each detector scheme, readout option, and gauge.
As an example, for electric field sensing in the magnetomechanical detector scheme the equation takes the gauge-independent form 
\begin{equation}
\frac{d \hat{a}}{dt} = i \Delta \hat{a} + i G_Q^{(E,v)} \hat{Q} - \sqrt{\kappa} \hat{b}_{\text{in}} - \frac{\kappa}{2} \hat{a} \enspace .
\end{equation}
Taking the difference between the equations which reference an initial time $t>t_0$ (in terms of $\hat{b}_{\text{in}}$) or a final time $t<t_0$ (in terms of $\hat{b}_{\text{out}}$) yields the familiar input-output relation
\begin{equation}
\label{eq:inout}
\hat{b}_{\text{out}}(t) = \hat{b}_{\text{in}}(t) + \sqrt{\kappa}\hat{a}(t) \enspace ,
\end{equation}
which describes how the output bath modes are related to the input bath modes and the cavity operator.

\begin{widetext}
\section{Explicit solutions for the output phase quadratures}
\label{noisePSD}

Here we list the solutions for the output phase quadratures found from different transducer and receiver combinations, as outlined in Section~\ref{S3} of the main text. For the magnetomechanical detector configuration, voltage measurement via electric field sensing yields the solution
\begin{equation}
\begin{split}
\hat{Y}_{\text{out}}^{(E,v)}(\nu) = &-\left( \frac{\frac{\kappa}{2} - i \nu}{\frac{\kappa}{2} + i \nu} \right) \hat{Y}_{\text{in}} + \frac{\hbar {G_Q^{(E,v)}}^2 \kappa (\nu^2 - \omega_m^2) }{L (\frac{\kappa}{2} + i \nu)^2 \left[  (\nu^2 - \omega_c^2)(\nu^2 - \omega_m^2) - \delta_v^2 \nu^2  \right]} \hat{X}_{\text{in}} \\
&+ \frac{i G_Q^{(E,v)} \sqrt{\kappa} \delta_v  \nu}{ \sqrt{m L}(\frac{\kappa}{2} + i \nu) \left[  (\nu^2 - \omega_c^2)(\nu^2 - \omega_m^2) - \delta_v^2 \nu^2  \right]} \hat{F}_{\text{in}}
\end{split} \enspace ,
\end{equation}
while for current measurement via magnetic field sensing we find the expression
\begin{equation}
\begin{split}
\hat{Y}_{\text{out}}^{(B,v)}(\nu) = &-\left( \frac{\frac{\kappa}{2} - i \nu}{\frac{\kappa}{2} + i \nu} \right) \hat{Y}_{\text{in}} 
+ \frac{\hbar {G^B}^2 \kappa L \omega_c^2 \Big(\nu^2 - \omega_m^2 - \delta_v^2 \Big)}{(\frac{\kappa}{2} + i \nu)^2 \left[  (\nu^2 - \omega_c^2 - \omega_l^2)(\nu^2 - \omega_m^2) - \delta_v^2 ( \nu^2 - \omega_l^2)  \right]} \hat{X}_{\text{in}} \\
&+ \frac{G^B  \sqrt{\kappa L} \delta_v \omega_c^2}{\sqrt{m}(\frac{\kappa}{2} + i \nu) \left[  (\nu^2 - \omega_c^2 - \omega_l^2)(\nu^2 - \omega_m^2) - \delta_v^2 ( \nu^2 - \omega_l^2)  \right]} \hat{F}_{\text{in}}
\end{split} \enspace .
\end{equation}
For the electromechanical case, voltage measurement via electric field sensing yields 
\begin{equation}
\begin{split}
\hat{Y}_{\text{out}}^{(E,x)}(\nu) = &-\left( \frac{\frac{\kappa}{2} - i \nu}{\frac{\kappa}{2} + i \nu} \right) \hat{Y}_{\text{in}} + \frac{\hbar \kappa \left[  {G_Q^{(E,x)}}^2 m (\nu^2 - \omega_m^2) - 2 G_Q^{(E,x)} G_x \sqrt{m L} \delta_x \omega_{\rm ce} + G_x^2 L (\nu^2 - \omega_{\rm ce}^2) \right]}{m L (\frac{\kappa}{2} + i \nu)^2 \left[  (\nu^2 - \omega_{\rm ce}^2)(\nu^2 - \omega_m^2) - \delta_x^2 \omega_{\rm ce}^2  \right]} \hat{X}_{\text{in}} \\
&+ \frac{ \kappa \left[  G_x L (\nu^2 - \omega_{\rm ce}^2) -  G_Q^{(E,x)} \sqrt{m L} \delta_x \omega_{\rm ce} \right]}{m L (\frac{\kappa}{2} + i \nu) \left[  (\nu^2 - \omega_{\rm ce}^2)(\nu^2 - \omega_m^2) - \delta_x^2 \omega_{\rm ce}^2  \right]} \hat{F}_{\text{in}}
\end{split} \enspace ,
\end{equation}
while for current measurement via magnetic field sensing we find the expression
\begin{equation}
\begin{split}
\hat{Y}_{\text{out}}^{(B,x)}(\nu) = &-\left( \frac{\frac{\kappa}{2} - i \nu}{\frac{\kappa}{2} + i \nu} \right) \hat{Y}_{\text{in}} 
+ \frac{\hbar {G^B}^2 \kappa L \omega_{\rm ce}^2 \Big(\nu^2 - \omega_m^2 + \delta_x^2 \Big)}{(\frac{\kappa}{2} + i \nu)^2 \left[  (\nu^2 - \omega_{\rm ce}^2 - \omega_{\rm le}^2)(\nu^2 - \omega_m^2) - \delta_x^2 ( \omega_{\rm ce}^2 + \omega_{\rm le}^2)  \right]} \hat{X}_{\text{in}} \\
&+ \frac{ i G^B T_x \sqrt{\kappa L} \omega_{\rm ce} \nu}{\sqrt{m}(\frac{\kappa}{2} + i \nu) \left[  (\nu^2 - \omega_{\rm ce}^2 - \omega_{\rm le}^2)(\nu^2 - \omega_m^2) - \delta_x^2 ( \omega_{\rm ce}^2 + \omega_{\rm le}^2)  \right]} \hat{F}_{\text{in}}
\end{split} \enspace .
\end{equation}
In all instances, we have utilized the frequencies defined Eq.~(\ref{eq:scales}) of the main text.

\section{Explicit noise PSD solutions}
\label{Full PSD}

In this appendix, we list the noise PSD expressions for each combination of detector configuration and measurement scheme. In contrast to those shown in the main text, here we write these expressions explicitly in terms of their frequency dependence and the relevant frequencies given in Eq.~(\ref{eq:scales}). We also include the coupling $G_x$ present in the electromechanical scheme, which we neglect due to its small size in the expressions in the main text.
For the magnetomechanical detector configuration, voltage measurement via electric field sensing yields the noise~PSD
\begin{equation}
\centering
S_{\text{FF}}^{(E,v)} = \frac{ {\hbar^2 G_Q^{(E,v)}}^2 \kappa m (\nu^2 - \omega_m^2)^2}{2 L \delta_v^2 \nu^2 \big(\frac{\kappa^2}{4}+ \nu^2 \big)} 
+ \frac{m L \big(\frac{\kappa^2}{4}+ \nu^2 \big) \left[ (\nu^2 - \omega_m^2) (\nu^2 - \omega_c^2) - \delta_v^2 \nu^2 \right]^2 }{ 2  {G_Q^{(E,v)}}^2 \kappa \delta_v^2  \nu^2} +   N_{\rm BM} \enspace ,
\end{equation}
while for current measurement via magnetic field sensing we find the expression
\begin{equation}
\centering
S_{\text{FF}}^{(B,v)} = \frac{\hbar^2 {G^B}^2  \kappa m L \Big( \nu^2 - \omega_m^2 - \delta_v^2 \Big)^2}{2 \delta_v^2 \big(\frac{\kappa^2}{4}+ \nu^2 \big)} + \frac{m \big(\frac{\kappa^2}{4}+ \nu^2 \big) \left[ (\nu^2 - \omega_m^2) (\nu^2 - \omega_c^2 - \omega_l^2) - \delta_v^2 (\nu^2 - \omega_l^2) \right]^2}{2  {G^B}^2 \kappa L \delta_v^2  \omega_c^4} +    N_{\rm BM} \enspace .
\end{equation}
In the electromechanical case, voltage measurement via electric field sensing yields the noise PSD
\begin{equation}
\centering
\begin{split}
S_{\text{FF}}^{(E,x)} = &\frac{ \hbar^2 \kappa \left[ {G_Q^{(E,x)}}^2 m  (\nu^2 -\omega_m^2) - 2 G_Q^{(E,x)} G_x \sqrt{m L} \delta_x \omega_{\rm ce} + G_x^2 L (\nu^2 - \omega_{\rm ce}^2) \right]^2}{2 \big( \frac{\kappa^2}{4}+ \nu^2 \big)  \left[ G_Q^{(E,x)} \sqrt{m L} \delta_x \omega_{\rm ce} -  G_x L (\nu^2 -\omega_{\rm ce}^2) \right]^2} \\
&+ \frac{ m^2 L^2 \big( \frac{\kappa^2}{4}+ \nu^2 \big) \left[ (\nu^2 - \omega_m^2) (\nu^2 - \omega_{\rm ce}^2) - \delta_x^2 \omega_{\rm ce}^2 \right]^2 }{ 2 \kappa  \left[ G_Q^{(E,x)} \sqrt{m L} \delta_x \omega_{\rm ce} -  G_x L (\nu^2 -\omega_{\rm ce}^2) \right]^2} + N_{\rm BM} \enspace , 
\end{split}
\end{equation}
while for current measurement via magnetic field sensing we find the expression
\begin{equation}
\centering
S_{\text{FF}}^{(B,x)} = \frac{\hbar^2 {G^B}^2 \kappa m L \omega_{\rm ce}^2 \Big( \nu^2 - \omega_m^2 + \delta_x^2 \Big)^2}{2 \delta_x^2 \nu^2 \big( \frac{\kappa^2}{4}+ \nu^2 \big)} + \frac{ m \big( \frac{\kappa^2}{4}+ \nu^2 \big) \left[ (\nu^2 - \omega_m^2) (\nu^2 - \omega_{\rm ce}^2 - \omega_{\rm le}^2) - \delta_x^2 (\omega_{\rm ce}^2 + \omega_{\rm le}^2) \right]^2 }{ 2 {G^B}^2 \kappa  L \delta_x^2 \omega_{\rm ce}^2 \nu^2} +  N_{\rm BM} \enspace .
\end{equation}
\pagebreak
\end{widetext}

\section{Analysis of optomechanical systems for comparison}
\label{Previous-Work}

Here we consider the continuous measurement of an optomechanical system subject to either direct position or direct momentum coupling. In the case where the probing optical amplitude quadrature $\hat{X}$ directly interacts with the position $\hat{x}$ of the mechanical system, the interaction Hamiltonian takes the form
\begin{equation}
\hat{H}_{\rm int} = \hbar G  \hat{x} \hat{X} \enspace .
\end{equation}
This is the basis of the standard optomechanical position-sensing problem. If instead the optical quadrature directly interacts with the velocity, i.e., the mechanical momentum of the system, which can be practically implemented by specific designs of the system as in Refs.~\cite{sohitri, ghosh2022combining}, the interaction Hamiltonian becomes
\begin{equation}
\hat{H}_{\rm int}= \hbar  G^\prime \hat{p} \hat{X} \enspace .
\end{equation}

Following the standard procedures in Appendix~\ref{IOtheory} and Section~\ref{S3}, we write down the full Hamiltonian and derive the equations of motion for these systems. Here we additionally consider a mechanical damping with damping rate $\mu$.
Then, we solve for the output phase quadrature of light $\hat{Y}_{\text{out}}$ using the input-output relations. These yield the estimated force expressions \cite{ghosh2022combining}
\begin{equation}
\begin{split}
\hat{F}_{E_x} 
&= - G \hbar \chi_c  \hat{X}_{\rm in} + \frac{e^{i \phi_c }\hat{Y}_{\rm in}}{ G\chi_c \chi_m}  + \hat{F}_{\rm in} \enspace ,
\end{split}
\end{equation}
and
\begin{equation}
\begin{split}
\hat{F}_{E_v} 
&=- i G^\prime  \hbar \chi_c m \frac{\omega_m^2}{\nu} \hat{X}_{\rm in} + \frac{i e^{i \phi_c}}{ G^\prime m \nu \chi_c \chi_m} \hat{Y}_{\rm in} + \hat{F}_{\rm in} \enspace .
\end{split}
\end{equation}
Here we define the cavity response function $\chi_c$, the mechanical response function $\chi_m$, and the cavity phase shift~$e^{i \phi_c}$ as
\begin{equation}
\begin{split}
\chi_c &= \frac{\sqrt{\kappa}}{-i \nu+\kappa/2} \enspace ,\\ 
\chi_m &= \frac{-1}{m(\nu^2-\omega_m^2 +i\mu \nu)} \enspace ,\\
e^{i \phi_c} &= \frac{-i \nu - \kappa/2}{-i \nu + \kappa/2} \enspace .
\end{split}
\end{equation}
For position sensing, this helps us establish the target frequency-dependent optimization of the coupling strength $G$: 
\begin{equation}
G(\nu) \rightarrow \frac{1}{\sqrt{\hbar} |\chi_m(\nu)|^{1/2} |\chi_c(\nu)|} \enspace .
\label{Optimized G}
\end{equation}

Using these estimated force expressions, we derive the noise PSD solutions analogously to those derived in the main text and use them to generate the broadband noise PSD plot shown Fig.~\ref{previous} for comparison with that of our electrical readout schemes.

\bibliography{IRrefs}

\end{document}